%% file: main.tex
\documentclass[manuscript]{acmart}
\AtBeginDocument{%
  \providecommand\BibTeX{{%
    Bib\TeX}}}

\usepackage{amsfonts}
\usepackage{algorithmic}
\usepackage{textcomp}
\usepackage{xcolor}
\def\BibTeX{{\rm B\kern-.05em{\sc i\kern-.025em b}\kern-.08em
    T\kern-.1667em\lower.7ex\hbox{E}\kern-.125emX}}

\newcommand{\tool}{$AiRacleX$}
\usepackage[autostyle=false, style=english]{csquotes}
\MakeOuterQuote{"}
         
\usepackage{enumitem} 
\usepackage{hyperref}

\usepackage{tcolorbox}

\usepackage[ruled,linesnumbered]{algorithm2e}
\usepackage{semantic}
\usepackage{tabularx}
\usepackage{array}

\usepackage{adjustbox}
\usepackage{multirow}

\input{chapters/solidity-highlight}

\begin{document}

\title{\tool:~ Automated Detection of Price Oracle Manipulations via LLM-Driven Knowledge Mining and Prompt Generation}

\author{Bo Gao}
\email{gao\_bo@ihpc.a-star.edu.sg}
\orcid{0000-0002-9812-521X}
\author{Yuan Wang}
\author{Qingsong Wei}
\author{Yong Liu}
\author{Rick Siow Mong Goh}
\affiliation{%
  \institution{Institute of High Performance Computing (IHPC), Agency for Science, Technology and Research (A*STAR)}
  \city{Singapore}
  \country{Singapore}
}

\author{David Lo}
\affiliation{%
  \institution{Singapore Management University}
  \city{Singapore}
  \country{Singapore}
}


\begin{abstract}
Decentralized finance (DeFi) applications depend on accurate price oracles to ensure secure transactions, yet these oracles are highly vulnerable to manipulation, enabling attackers to exploit smart contract vulnerabilities for unfair asset valuation and financial gain.
Detecting such manipulations traditionally relies on the manual effort of experienced experts, presenting significant challenges.

In this paper, we propose a novel LLM-driven framework that automates the detection of price oracle manipulations by leveraging the complementary strengths of different LLM models (LLMs).
Our approach begins with domain-specific knowledge extraction, where an LLM model synthesizes precise insights about price oracle vulnerabilities from top-tier academic papers, eliminating the need for profound expertise from developers or auditors. This knowledge forms the foundation for a second LLM model to generate structured, context-aware chain of thought prompts, which guide a third LLM model in accurately identifying manipulation patterns in smart contracts. 
We validate the framework’s effectiveness through experiments on 60 known vulnerabilities from 46 real-world DeFi attacks or projects spanning 2021 to 2023.
The best performing combination of LLMs (Haiku-Haiku-4o-mini) identified by \tool~demonstrate a 2.58-times improvement in recall (0.667 vs 0.259) compared to the state-of-the-art tool GPTScan, while maintaining comparable precision.
Furthermore, our framework demonstrates the feasibility of replacing commercial models with open-source alternatives, enhancing privacy and security for developers.


\end{abstract}




\maketitle

\input{chapters/1_introduction}
\input{chapters/2_background}

\input{chapters/3_method}

\input{chapters/4_evaluation}
\input{chapters/5_related}

\input{chapters/6_conclusion}

\bibliographystyle{ACM-Reference-Format}


\input{chapters/appendix}

\end{document}

%% file: chapters/solidity-highlight.tex
\usepackage{listings, xcolor, lstautogobble}

\definecolor{verylightgray}{rgb}{.97,.97,.97}

\lstdefinelanguage{Solidity}{
	keywords=[1]{anonymous, assembly, assert, balance, break, call, callcode, case, catch, class, constant, continue, contract, debugger, default, delegatecall, delete, do, else, event, export, external, false, finally, for, function, gas, if, implements, import, in, indexed, instanceof, interface, internal, is, length, library, log0, log1, log2, log3, log4, memory, modifier, new, payable, pragma, private, protected, public, pure, push, require, return, returns, revert, selfdestruct, send, storage, struct, suicide, super, switch, then, this, throw, transfer, true, try, typeof, using, value, view, while, with, addmod, ecrecover, keccak256, mulmod, ripemd160, sha256, sha3}, 
	keywordstyle=[1]\color{blue}\bfseries,
	keywords=[2]{address, bool, byte, bytes, bytes1, bytes2, bytes3, bytes4, bytes5, bytes6, bytes7, bytes8, bytes9, bytes10, bytes11, bytes12, bytes13, bytes14, bytes15, bytes16, bytes17, bytes18, bytes19, bytes20, bytes21, bytes22, bytes23, bytes24, bytes25, bytes26, bytes27, bytes28, bytes29, bytes30, bytes31, bytes32, enum, int, int8, int16, int24, int32, int40, int48, int56, int64, int72, int80, int88, int96, int104, int112, int120, int128, int136, int144, int152, int160, int168, int176, int184, int192, int200, int208, int216, int224, int232, int240, int248, int256, mapping, string, uint, uint8, uint16, uint24, uint32, uint40, uint48, uint56, uint64, uint72, uint80, uint88, uint96, uint104, uint112, uint120, uint128, uint136, uint144, uint152, uint160, uint168, uint176, uint184, uint192, uint200, uint208, uint216, uint224, uint232, uint240, uint248, uint256, var, void, ether, finney, szabo, wei, days, hours, minutes, seconds, weeks, years},	
	keywordstyle=[2]\color{teal}\bfseries,
	keywords=[3]{block, blockhash, coinbase, difficulty, gaslimit, number, timestamp, msg, data, gas, sender, sig, value, now, tx, gasprice, origin},	
	keywordstyle=[3]\color{violet}\bfseries,
	identifierstyle=\color{black},
	sensitive=false,
	comment=[l]{//},
	morecomment=[s]{/*}{*/},
	commentstyle=\color{gray}\ttfamily,
	stringstyle=\color{red}\ttfamily,
	morestring=[b]',
	morestring=[b]"
}

%% file: chapters/1_introduction.tex
\section{Introduction} \label{sec:introduction}
Decentralized finance (DeFi) has emerged as a groundbreaking paradigm, revolutionizing the landscape of traditional finance by offering open, accessible, and permissionless financial services built on the foundation of blockchain technology.
At the core of many DeFi applications lie price oracles, providing essential external price data for smart contracts that power a wide array of financial activities, including lending, borrowing, trading, and more.
By delivering accurate up-to-date price information, price oracles enable DeFi protocols to operate effectively, execute fair transactions, and maintain ecosystem stability.
However, the critical role of price oracles also makes them a prime target for exploitation, posing a significant risk to the integrity of the entire ecosystem.
Manipulating the price data provided by an oracle to mislead smart contracts about the true value of an asset is referred to \emph{price oracle manipulation (POM)}~\cite{wu2021defiranger}.
Through POM, adversaries can exploit misled smart contracts to gain unfair advantages or illicit profits.
These attacks can take various forms, such as using flash loans to temporarily distort asset prices, compromise data sources to feed false information to oracles, influence decentralized voting mechanisms to distort price data, or exploit time-weighted average price (TWAP) calculations to create inaccurate price feeds.
Recent studies have highlighted the severity of the issue. For example, Zhou et al.~\cite{zhou2023sok} reported that on-chain oracle manipulation incidents are the most common protocol layer incident type, acconting for 15\% of the total real-world attacks analyzed.
Similarly, Zhang et al.\cite{zhang2023demystifying} find that price oracle manipulation exploits represent 34.3\% of the exploits in their real-world dataset, identifying it as the most common exploit among machine-unauditable bugs (MUBs)—a category of vulnerabilities that, as of 2022, remain undetectable by existing automated tools. 

Although recent years have seen advancements in automated security tools, the blockchain community still struggles to effectively address POM. 
Chaliasos et al.~\cite{chaliasos2024smart} highlight that only 25\% of real-world attacks are detected by widely used static analysis tools such as ConFuzzius~\cite{torres2021confuzzius}, Mythril~\cite{consensys2024mythril}, Oyente~\cite{luu2016oyente}, Slither~\cite{feist2019slither}, and Solhint~\cite{protofire2024solhint}, underscoring the persistent limitations of existing automated approaches.
Few works attempting to address POM include DeFiRanger~\cite{wu2021defiranger}, which identifies price oracle manipulation attacks using pattern matching; ProMutator~\cite{wang2021promutator}, which simulates potential price manipulation attacks to identify weak points in oracle systems; DeFiPoser~\cite{zhou2021defiposer}, which uses SMT solvers to detect complex profitable transactions; and OVer~\cite{deng2024safeguarding}, which employs symbolic analysis and SMT solvers to ensure the secure operation of DeFi protocols.
These approaches often require extensive transaction data, significant computational resources, or accurate modeling, limiting their performance and adoption.
In practice, this kind of task still heavily relies on the manual efforts of experienced experts by analyzing data and patterns.
Thus, these approaches are inherently limited by human capacity and expertise, making it difficult to scale and adapt to the rapidly evolving landscape of DeFi and smart contract technologies.

With the rapid development of large language models (LLMs), some works have leveraged LLMs for detecting smart contract vulnerabilities.
Early exploration by Issac et al.~\cite{david2023you} demonstrated the effectiveness of ChatGPT-4 and Claude in conducting smart contract security audits especially identifying logic issues and coding errors, but they generated a significant number of false positives (95\% of alarms).
Building upon this, Gao et al.\cite{gao2024unveiling} further evaluated ChatGPT-4's performance across six specific categories of bugs, revealing that it can detect an average of 15\% of vulnerabilities using bug-type-agnostic prompts. Notably, while it successfully identified 33\% of price oracle manipulation bugs, this came with an 87\% false positive rate, highlighting the complex trade-off between recall and precision in LLM-based approaches.
These initial efforts use LLMs in a straightforward manner and are more evaluation-oriented, lacking dedicated design for POM.
Following these works, GPTLens~\cite{hu2023large} introduced an adversarial framework with LLMs serving as an \textit{Auditor} and \textit{Critic} to improve vulnerability detection. By separating generation and discrimination stages, GPTLens reduces false positives while maintaining recall. However, its models struggle to interpret ground truth effectively, resulting in only marginal improvements in a small evaluation of 13 projects.
Notably, GPTScan~\cite{sun2023gptscan} combines ChatGPT’s code analysis capabilities with static analysis to detect logical vulnerabilities in smart contracts. While it achieves high precision and recall on detecting vulnerability types, it struggles with complex vulnerabilities like POM, which require long and intricate function calls. GPTScan also depends heavily on ChatGPT's output format, which is prone to errors even in streamlined JSON formats, complicating integration with static analysis tools. This limitation leads to missed vulnerabilities or false alarms when validating strategies by static analysis tools are loosened.
Our evaluation confirmed these challenges, showing increased false positive and false negative rates when GPTScan's results are assessed with standard metrics. Moreover, its fixed warning messages for each vulnerability type limit interpretability, and the static analysis tools it relies on often fail to handle diverse smart contracts.
These studies underscore LLMs’ potential in advancing smart contract security but highlight persistent challenges, including high false positive rates, limited coverage of complex bugs, and difficulty delivering actionable insights for real-world applications.
To overcome the limitations of static analysis tools, our work introduces a novel framework leveraging three LLM models, designated as \textit{Knowledge Synthesizer}, \textit{Prompt Generator} and \textit{Auditor}.
The process begins with the Knowledge Synthesizer, an LLM dedicated to the extraction and synthesis of domain-specific insights from top-tier academic literature. 
This initial step is essential, as it filters out noise introduced by lower-quality data sources—such as online forums, blogs, and miscellaneous webpages—commonly found in training datasets. Moreover, it supplies the pipeline with precise, externally validated domain knowledge.
Building on the high-fidelity insights provided by the Knowledge Synthesizer, the Prompt Generator plays a critical role in translating these insights into structured and actionable chain of thought (CoT) prompts. 
This method, demonstrated to significantly enhance LLM performance in various applications~\cite{kojima2022large}, ensures that the Auditor is guided with precise and contextually relevant instructions.
Equipped with these tailored prompts, the Auditor enhances the detection of POM across diverse projects.
To evaluate the effectiveness of our framework, we conducted experiments on a dataset containing 36 bugs from 31 real-world DeFi attacks between 2021 and 2022, as well as 24 bugs from 15 Code4rena projects spanning 2021 to 2023. 
Our results demonstrate a 2.58-times improvement in recall (0.667 vs 0.259) compared to the state-of-the-art tool GPTScan, while maintaining comparable precision. 
Additionally, when compared to the zero-shot CoT prompt, our approach achieves a 15\% increase in precision (0.313 vs 0.271) with comparable recall.
Moreover, our approach streamlines the process for developers, eliminating the need for domain knowledge in smart contracts or crafting problem-specific prompts.
To conclude, we make the following contributions:
\begin{itemize}


\item A Novel and Transferable Multi-LLM Framework:
We propose a multi-LLM framework that synergistically integrates a domain-specific \textit{Knowledge Synthesizer}, an optimized \textit{Prompt Generator}, and an automated \textit{Auditor} to effectively identify POM vulnerabilities in smart contracts. 
The \textit{Knowledge Synthesizer} enables seamless extension to other vulnerabilities without the need for predefined rules or code modifications, as seen in tools like GPTScan, by minimizing reliance on expert knowledge.
Meanwhile, the \textit{Prompt Generator} facilitates the automatic creation of structured prompts, eliminating the need for manual intervention and enhancing efficiency.

\item Optimized Model Selection and Knowledge Evolution:
We evaluate and identify the optimal combination of LLM models for knowledge summarization, prompt generation, and vulnerability detection. Our results highlight the complementary strengths of different models, showcasing the effectiveness of manually curated knowledge in improving detection capabilities. Furthermore, we demonstrate how LLM-based knowledge synthesizers can replicate and surpass human-curated performance, paving the way for fully automated systems.

\item Comprehensive Validation and New Discoveries:
Through extensive evaluation on diverse real-world datasets, including historical DeFi exploits and Code4rena projects, we validate the effectiveness of our framework. 
Notably, our approach successfully identified $20$ bugs which can only be detected by \tool~but not by SOTA tool GPTScan.
\end{itemize}

\paragraph{Outline} In the subsequent sections of this paper, we introduce some essential concept in Section~\ref{sec:background}. Then, we delve deeper into the methodology of our approach, outlining the process of knowledge extraction, prompt generation and automatic audit in Section~\ref{sec:method}.
In Section~\ref{sec:evaluation}, we present the results of our experiments and discuss the implications of our findings for the broader blockchain and DeFi ecosystems. Section~\ref{sec:related} delves into related works and we conclude our paper in Section~\ref{sec:conclusion}.

%% file: chapters/2_background.tex
\section{Preliminaries} \label{sec:background}
This section provides some basics about POM, including how price oracle manipulation occurs, the typical POMs and the representative causes of POM vulnerabilities.
We assume some familiarity with basic concepts such as blockchain, Ethereum, and smart contracts, and refer readers to~\cite{wood2014ethereum} for details. 

\subsection{Types of DeFi Applications}
DeFi applications aim to provide financial services without traditional intermediaries, leveraging blockchain technology and smart contracts.
The main types of DeFi applications include decentralized exchanges (DEXs), stablecoins, lending and borrowing platforms, yield farming and liquidity mining, and decentralized autonomous organizations (DAOs) etc.
This work further classifies DeFi applications into two categories based on their role in the ecosystem:
\paragraph{Price Provider Applications} These applications provide price data to other DeFi applications. Examples include Chainlink oracle contracts, decentralized exchanges (DEXs), and other protocols that define their own logic for determining asset prices.
\paragraph{Price Consumer Applications} Applications falling under this category rely on accurate price data for their operational efficacy. Examples include lending and borrowing platforms, which use price data to determine the valuation of collaterals, and DAOs, which use price data to determine voting weights and other governance parameters.

With this classification, we can accurately analyze the root causes of the price oracle vulnerabilities and design efficient prompts to help LLMs detect such kinds of vulnerabilities.

\subsection{Types of Price Oracles}
Price oracles can be generally categorized into three classes: on-chain oracles, off-chain oracles, and hybrid price oracles. 
    \paragraph{On-Chain Oracles} On-chain oracles, like those used by Uniswap, derive price data directly from on-chain activities such as trading within liquidity pools. These oracles use mechanisms like Constant Product Formula (CPF)~\ref{uniswapcpf} to calculate asset prices based on real-time transactions occurring on the blockchain. Despite their susceptibility to manipulation in low-liquidity scenarios, they offer several advantages:
    \begin{itemize}
        \item Native to Blockchain: Since on-chain oracles operate entirely within the blockchain environment, they provide seamless integration with decentralized applications (DApps) and smart contracts without the need for external dependencies.
        \item Real-time Pricing: Prices reflect current market conditions as they are derived directly from ongoing transactions on the blockchain.
        \item Decentralization: Since these oracles are based on decentralized mechanisms (e.g., Uniswap’s liquidity pools), there is no central authority controlling the price feed, reducing single points of failure.
        \item Full Transparency: Anyone can verify the price data on-chain, ensuring the data's integrity and preventing manipulation by a central entity.
    \end{itemize}

    \paragraph{Off-Chain Oracles} Off-chain oracles, exemplified by Chainlink~\footnote{https://chain.link/}, gather data from external sources and bring it onto the blockchain through a decentralized network of node operators. Their advantages can be summarized as follows:
    
    \begin{itemize}
        \item Access to Diverse Data: Off-chain oracles can pull price information from a wide range of external sources, including traditional financial markets, making them suitable for use cases requiring data beyond the blockchain ecosystem.
        \item Robustness to Manipulation: Since off-chain oracles aggregate data from multiple independent nodes or sources, they are generally more resistant to manipulation or data skewing compared to purely on-chain systems.
        \item Scalability: They are often more scalable than on-chain oracles since they are not dependent on the blockchain’s transaction throughput and can aggregate large volumes of data from diverse sources without congesting the network.
    \end{itemize}
    
    \paragraph{Hybrid Oracles} Hybrid oracles, such as those used by Extra Finace\footnote{https://docs.extrafi.io/extra\_finance/leverage-farming/price-feed} combine features of both on-chain and off-chain oracles to enhance price stability and security. Typically, the prices are derived directly from on-chain data provided by decentralized exchanges (DEXs). 
    However, to mitigate the risk of abnormal price fluctuations, Chainlink price feeds are employed as a safeguard. 
    This hybrid approach offers several advantages:

    \begin{itemize}
        \item Comprehensive Data Validation: Hybrid oracles cross-reference off-chain and on-chain data to ensure both accuracy and consistency, reducing the risk of manipulation.
        \item Real-Time Responsiveness: The integration of on-chain mechanisms ensures timely updates to price data, even during volatile market conditions.
        \item Resilience to Attacks: The use of diverse data sources creates redundancy, making hybrid oracles more robust against single-point failures or targeted attacks.
    \end{itemize}

    Despite the popularity of \textit{Off-Chain Oracles} like Chainlink, \textit{On-Chain Oracles} remain a viable option for many in the blockchain community. 
    According to an oracle dashboards \footnote{https://defillama.com/oracles} \footnote{https://defillama.com/oracles/TWAP}, as many as 90 projects opted for On-Chain Oracles.
    This preference is largely due to their inherent consistency with the decentralization philosophy of blockchain systems, as they operate entirely within the blockchain environment, ensuring trustlessness and minimizing reliance on external entities.

\subsection{Price Oracles Manipulation (POM)}
POM can stem from various sources based on above types of oracles. The primary complications arise from on-chain and off-chain oracles.
\paragraph{On-Chain Price Oracle Manipulation}
On-chain oracles can be easily manipulated due to their reliance on spot prices from a single source. For instance, an attacker can use a flash loan to temporarily drain liquidity from a pool, causing the price to be artificially inflated or deflated. This manipulation allows the attacker to exploit the manipulated price, leading to significant financial gains, as demonstrated in the PancakeBunny attack\footnote{https://medium.com/amber-group/bsc-flash-loan-attack-pancakebunny-3361b6d814fd}.
A more detailed example is illustrated in the Appendix~\ref{app:oraclemanip}.

\paragraph{Off-Chain Price Oracle Manipulation}
Off-chain oracles face different challenges. Centralized off-chain oracles depend on a single trusted entity, making them vulnerable to malicious data submission by authorized users for personal gain. Additionally, the compromise of private keys can pose significant risks. Decentralized off-chain oracles mitigate some of these risks by aggregating data from multiple sources, but they are not immune to issues like freeloading or Sybil attacks among data collectors. Further, off-chain infrastructure vulnerabilities—including those in access control, cryptographic implementations, transport, and database security—add layers of complexity in preventing manipulation~\cite{offchainoracle}.

While these issues are broad and affect the overall security of price oracles, this paper focuses specifically on vulnerabilities that adversaries can exploit, particularly through specialized inputs to on-chain contracts. This includes manipulations involving on-chain price oracles and the on-chain components of off-chain price oracles, which can lead to significant financial losses or unfair advantages for attackers.

\subsection{Causes of Price Oracle Manipulation}
Price oracle manipulation arises from various factors that exploit weaknesses in both the underlying mechanisms and the broader DeFi ecosystem. Below are some key causes:

\paragraph{Smart Contract Vulnerabilities} Careless bugs or flawed logic while development in the smart contracts governing liquidity pools or price feed mechanisms can lead to incorrect pricing, enabling attackers to manipulate asset values and potentially causing significant financial losses for users and protocols~\cite{gao2024unveiling}.

\paragraph{Flash Loan Attacks} Flash loans allow users to borrow large amounts of capital without collateral, provided the loan is repaid within the same transaction. Attackers exploit this feature by executing large trades to temporarily inflate or deflate the price of assets in on-chain liquidity pools. This manipulated price can then be leveraged in other DeFi protocols that depend on the oracle, leading to cascading financial consequences~\cite{zhang2023demystifying}.

\paragraph{Front-Running Attacks} Front-running attacks, enabled by the transparency of blockchain transactions, also contribute to price oracle manipulation. Malicious actors monitor pending transactions and strategically place their trades just before large transactions. By doing so, they can profit from the resulting price changes while distorting the price data in liquidity pools~\cite{zhang2023demystifying}.

\paragraph{Impermanent Loss Impact} Liquidity providers may suffer from impermanent loss, where the value of their deposited assets changes due to price fluctuations within the pool. If a DeFi application relies on the pool's price without accounting for these fluctuations, it might overestimate or underestimate the true value of assets~\cite{labadie2022impermanent}.

\paragraph{Slippage} The difference between the expected and actual executed price of a trade presents another avenue for manipulation. In low-liquidity pools, attackers can exploit slippage by executing large trades that cause significant price deviations. These deviations can propagate through dependent DeFi applications, leading to inaccurate price feeds and destabilizing the broader ecosystem~\cite{labadie2022impermanent}.

These factors are inherent features of blockchain and DeFi systems, not deficiencies. While they do not inherently lead to attacks, they can introduce vulnerabilities under certain conditions. The goal is not to eliminate these features but to identify potential weaknesses and mitigate their adverse effects, thereby maximizing their benefits.

%% file: chapters/3_method.tex
\section{Proposed Approach} \label{sec:method}
In this section, we first briefly review the state-of-the-art LLMs and key prompt engineering techniques, establishing the foundations of our work. We then introduce our LLM-driven detection framework, \tool, in detail.

\begin{figure}[h]
  \centering
  \includegraphics[width=\linewidth]{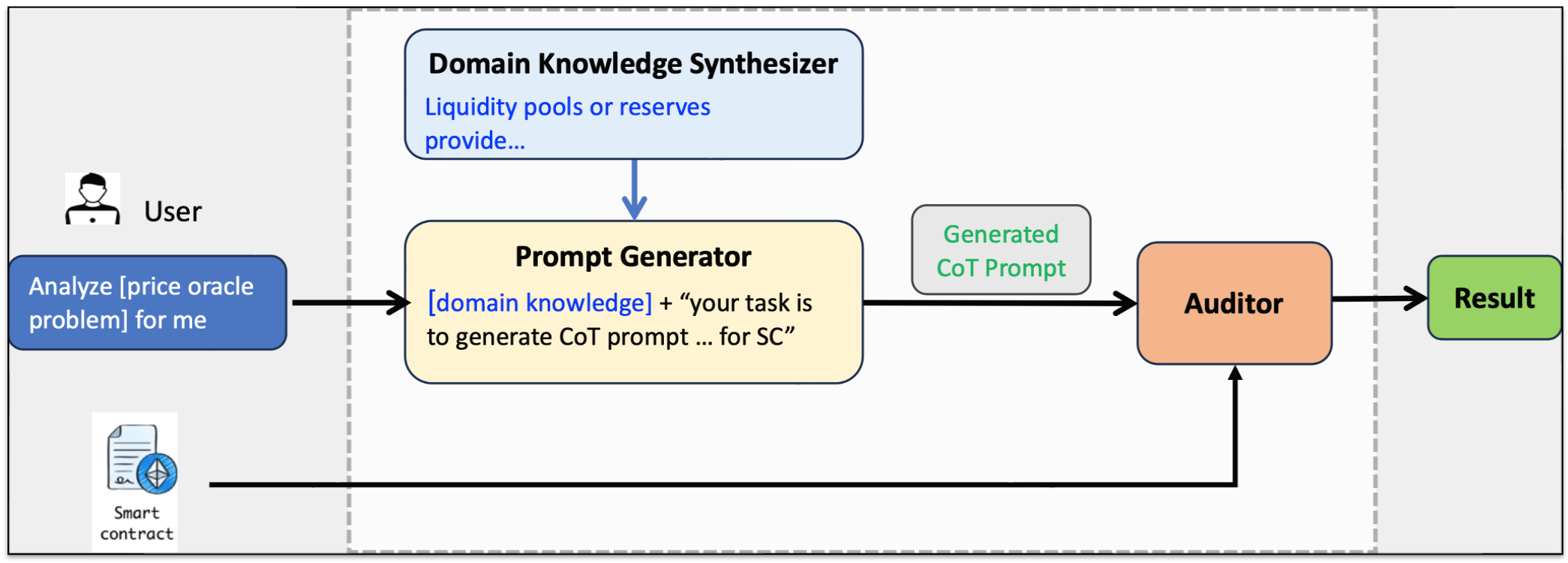}
  \caption{Overview of LLM-driven Automated Detection Framework.}
  \Description{The framework includes a knowledge synthesizer, a prompt generator and an auditor.}
  \label{fig:overview}
\end{figure}

\subsection{LLMs and Prompt Design Methods} \label{sec:promptmethod}
The rapid advancements in large language models (LLMs) have been driven by improvements in machine learning algorithms, computational power, and extensive training datasets.
State-of-the-art models like ChatGPT series\footnote{https://openai.com/chatgpt/overview}, Claude series\footnote{https://www.anthropic.com/claude}, and open-source models like Llama series\footnote{https://www.llama.com/}, Qwen series\footnote{https://www.alibabacloud.com/en/solutions/generative-ai/qwen}have significantly advanced natural language processing (NLP), excelling in tasks such as text generation, summarization, question answering, and program bug detection~\cite{li2023adapting}.
While LLMs have demonstrated remarkable performance across various tasks, their ability  in reasoning and addressing complex problems remains highly dependent on the quality of the prompts provided~\cite{wei2022chain}. 
To maximize their potential, innovative prompt engineering techniques have been developed, such as Chain-of-Thought (CoT)\cite{wei2022chain}, Least-to-Most\cite{zhou2022least}, and Complex CoT~\cite{fu2022complexity}, which guide models to decompose complex tasks into smaller, more structured steps. 
Remarkably, even simple zero-shot CoT prompts like "let's think step by step" have demonstrated improvements of up to 60\% on specific datasets~\cite{kojima2022large}.

Despite these advancements, designing efficient and effective prompts remains a challenge, particularly for complex tasks like POM detection. Building upon these techniques, our work integrates automated prompt design into a systematic, multi-LLM framework. By synthesizing domain-specific knowledge and dynamically generating tailored prompts, we enable LLMs to address the intricate challenge of detecting price oracle manipulation vulnerabilities in a scalable and automated manner. This eliminates reliance on manual intervention, achieving both precision and efficiency.


\subsection{\tool: LLM-driven Automated Detection Framework}
In this section, we introduce the three core components of \tool. As illustrated in Fig.~\ref{fig:overview}, our framework is composed of the \textbf{Domain Knowledge Synthesizer}, \textbf{Prompt Generator}, and \textbf{Auditor}, each playing a crucial role in automating and optimizing the detection process.

\subsubsection{Domain Knowledge Synthesizer}
This module synthesizes precise insights about price oracle vulnerabilities by leveraging top-tier academic literature, minimizing reliance on developer or auditor expertise through the automated summarization of key definitions and patterns.
Furthermore, it ensures the use of high-quality domain-specific information, mitigating the influence of web-scale noisy data used in the pre-training phase of the model.

\begin{tcolorbox}[float, colback=gray!20, colframe=gray!50, title={Knowledge Synthesizing Example}, label=knowledgesynthesizer]
\textbf{System:} You are a blockchain researcher. You are good at creating precise, comprehensive problem definitions. \\
\\
\textbf{User:} You are given descriptions of the price oracle manipulation problem from top academic conference papers, delimited by triple backticks \texttt{```}. Your task is to analyze these descriptions and provide a concise, unified definition of the price oracle manipulation problem. \texttt{```} + \{knowledge\} + \texttt{```}. Return the definition in the JSON format: \{output format\}. \\
\\
\textbf{Assistant:} The price oracle manipulation problem refers to a vulnerability in decentralized finance (DeFi) applications where an attacker can manipulate the price of a cryptocurrency in a decentralized exchange (DEX) that serves as the price oracle for the victim smart contract. The attack consists of two steps: 1. Price Oracle Manipulation (POM): The attacker trades a large volume of the cryptocurrency in the DEX, tilting the balance of the trading pair and causing the price to deviate from the actual market price. 2. Arbitrage: The attacker then exploits the price disparity between the manipulated price and the actual market price to obtain the cryptocurrency at the lower, manipulated price and sell it at the higher, actual market price, profiting from the difference. This vulnerability arises when the victim smart contract relies on the manipulated price provided by the DEX price oracle, rather than using more robust price oracles or mechanisms to mitigate price manipulation.
\end{tcolorbox}

To achieve this, we selected papers relevant to the topic of price oracle problems from prestigious conferences, such as the IEEE Symposium on Security and Privacy (S\&P), International Conference on Software Engineering (ICSE), IEEE European Symposium on Security and Privacy Workshops (EuroS\&PW), and ACM SIGSOFT International Symposium on Software Testing and Analysis (ISSTA).
From these sources, we identified seven papers~\cite{xi2024pomabuster,deng2024safeguarding,zhang2023demystifying,wu2021defiranger,wang2021promutator,kong2023defitainter, mo2023toward} that are particularly relevant to price oracle manipulation.
We manually excerpted paragraphs within these papers that describe price oracle problems, including definitions, common attack vectors, techniques, and defense mechanisms. 
We believe this manual step, while necessary, presents minimal challenges. This is because the published academic papers typically employ clear subtitles to delineate problem descriptions, making the identification and extraction of relevant information straightforward.

Subsequently, we prompt the LLM model to summarize the paragraphs on price oracle manipulation. This synthesized definition serves as the sole knowledge for the prompt generator, ensuring that the generated prompts are grounded in accurate and comprehensive domain knowledge.
An example of the knowledge synthesis process, including the prompt and response, is provided in \textit{Knowledge Synthesizing Example}~\ref{knowledgesynthesizer}.

This module is pivotal for extending our framework to address other issues. By extracting high-quality, relevant descriptions of specific problems, the framework can be easily adapted to tackle vulnerabilities such as \textit{Privilege Escalation} and \textit{Inconsistent State Updates}~\cite{zhang2023demystifying}.

\subsubsection{Prompt Generator and Auditor}


\begin{tcolorbox}[float, colback=gray!20, colframe=gray!50, title={CoT Prompt Generating Example}, label=cotprompt]
\textbf{System:} \textcolor{orange}{You are an experienced prompt engineer.} \\
\\
\textbf{User:} Your task is to provide a chain-of-thought prompt to an AI language model to accurately detect price oracle manipulation issues which could lead to financial losses or unfair advantages in the given smart contract \textcolor{purple}{step by step.} \textcolor{blue}{Remember to base your generated prompt solely on the price oracle description in triple backticks \texttt{```}}. \textcolor{red}{Do not make assumptions about external factors or systems not mentioned in the given information.} \texttt{```} + \textcolor{brown}{\{synthesized knowledge\}} + \texttt{```}. Return the prompt in the JSON format: \{output format\}. \\
\\
\textbf{Assistant:} description: Detect price oracle manipulation vulnerabilities in a smart contract,  steps:
step: 1, description: Identify the price oracle used by the smart contract. Determine if the price oracle is a decentralized exchange (DEX) or another external price feed.,     \\ 
questions: \\
\hspace*{1cm} a. What is the price oracle used by the smart contract?       \\
\hspace*{1cm} b. Is the price oracle a decentralized exchange (DEX) or another external price feed? \\
    step: 2, description: Analyze the smart contract's reliance on ...
\end{tcolorbox}

The synthesized domain knowledge is integrated into the \textbf{Prompt Generator}, a component that combines extracted insights with tailored instructions.
We employ several techniques to enhance the prompt generator's ability to produce structured and actionable CoT prompts.
As demonstrated in \textit{Prompt Generating Example}\ref{cotprompt}, the \textbf{System} and \textbf{User} represent the input provided by \tool, while the \textbf{Assistant} showcases the model’s output.
In this example, the sentence highlighted in orange represents a role-based prompt, which assigns a specific relevant role to guide the prompt. 
This technique has proven effective in various tasks, with Zhang et al.~\cite{zheng2023helpful} reporting approximately a 20\% improvement in accuracy compared to the \textit{Simple Prompt} approach.
Building on this, we introduced the \textit{Zero-shot CoT Prompt} approach, described in Section~\ref{sec:promptmethod}.
This method utilizes the "magical" phrase \textbf{step by step}, highlighted in brown, to encourage the model to perform logical reasoning.
Another technique employed is the use of positive and negative prompts, inspired by conditional generation models like Stable Diffusion~\cite{ban2025understanding}. 
Positive prompts, highlighted in blue, provide explicit guidance, while negative prompts, shown in red, define constraints to avoid irrelevant or misleading outputs.
These prompts are important because LLM models are pre-trained on general web-scale data, this pre-trained knowledge may conflict with the actual context of POM and thus interfere the analysis of the underlying task. For instance, such knowledge often leads to extraneous alarms, like presuming that an oracle owner’s potential to modify the oracle inherently makes the "set oracle" function vulnerable. 
While these findings may hold in broader contexts, they are not classified as POM attacks from a developer’s perspective and therefore increase false alarm (as revealed in the later section).
By adhering to the constraints defined by positive and negative prompts, the generated Cot prompt remain highly focused and relevant to the specific vulnerabilities under investigation.
Next, we incorporate synthesized domain knowledge, highlighted in brown, from the \textit{Knowledge Synthesizer} into the prompts, creating what we call \textit{Context-based Prompts}. 
With this enriched knowledge, the prompt generator can deliver task-aware and accurate instructions, ensuring that the prompts are highly aligned with the nuances of the vulnerabilities being analyzed. 
The complete prompt generated by this process is provided in Appendix~\ref{app:componentprompts}.

Lastly, the generated CoT prompt is utilized by the \textbf{Auditor} component, which employs an LLM model to detect POM vulnerabilities in the provided smart contract.
The input to the Auditor is shown in \textit{Auditor Prompt Example} below.

\begin{tcolorbox}[colback=gray!20, colframe=gray!50, title={Auditor Prompt Example}, label=prompt_auditor]
\textbf{System:} You are an experienced expert on auditing price oracle manipulation problems. Your task is to conduct a thorough audit on the provided solidity file to identify all potential price oracle manipulation vulnerabilities. \\

\textbf{User:} \{generated CoT prompt\} + \{smart contract\} + Respond the prompt in the following JSON format: \{output format\}.
\end{tcolorbox}

In addition to the techniques outlined above, we emphasize the importance of fostering internal reasoning before generating responses, aligning with OpenAI's \textbf{o1} reasoning model.\footnote{https://platform.openai.com/docs/guides/reasoning}
To minimize manual effort across components, we incorporate this concept into the output format.
As demonstrated in the \textit{Output Format} example, the model is instructed to include key fields such as \textbf{beneficiary}, \textbf{victim}, and \textbf{reason} for each finding. These fields encourage the model to engage in critical reasoning, ensuring that relevant actors are accurately identified and justifications are provided for each result.
This structured approach not only enhances the quality and interpretability of the output but also mitigates hallucination issues, reduces false alarms, and improves detection accuracy while minimizing noise in the analysis.

\begin{tcolorbox}[colback=gray!20, colframe=gray!50, title={Output Format Example}, label=outputformat]
    Analyze the smart contract delimited with \texttt{```}. Respond all the vulnerabilities with the following JSON format: \\
    \{vulnerable: yes, function: functionName, contract: contract name of the vulnerable function, beneficiary: ..., victim: ..., reason:...\} \\

    - vulnerable should be yes if the vulnerability exists, otherwise no. \\
    - \textbf{beneficiary} should be the role who will gain in the vulnerability. \\
    - \textbf{victim} should be the role who will suffer a loss or disadvantage in the vulnerability. \\
    - \textbf{reason} should describe why you think it is vulnerable and how to manipulate the price oracle to exploit this vulnerability.
\end{tcolorbox}

%% file: chapters/4_evaluation.tex
\section{Evaluation}\label{sec:evaluation}
In this section, we present the dataset utilized for our evaluation and discuss the findings derived from addressing the following research questions (RQs):

\begin{enumerate}[label=RQ\arabic*:, left=1em] 
\item How effective are the state-of-the-art (SOTA) tool GPTScan and a zero-shot CoT prompt approach?
\item How effective is \tool~when evaluated on diverse real-world projects with human-curated knowledge?
\item How does the Knowledge Synthesizer impact the framework's performance in comparison to human-curated knowledge?
\end{enumerate}

\subsection{Dataset} \label{dataset}
Our dataset is curated to reflect real-world scenarios and challenges, providing a robust benchmark for evaluating the framework. As shown in Table~\ref{table:dataset}, which summarizes the key statistics for the datasets, including the number of projects, the number of vulnerabilities for each dataset, and the average number of functions and lines of code (LoC) for each project, it generally consists of two categories: real-world attacked DeFiHacks projects and Code4Rena audit contest projects. To focus specifically on price oracle manipulation, we extended both datasets, ensuring comprehensive and reliable evaluation. 
By leveraging projects with documented exploitation reports and rigorous audit reports, we implemented a stringent method to categorize the vulnerabilities, thereby minimizing potential reporting biases and enhancing the empirical reliability of our evaluation framework. 

\begin{table}
    \caption{Datasets for Evaluation}
    \label{table:dataset}
    \begin{tabular}{@{}lcccc@{}}
    \toprule
    Dataset  & Projects & Vulnerabilities & Functions (Avg) & LoC (Avg)\\ \midrule
    DeFiHack  & 31       & 36      & 24  &  1,630 \\
    Code4Rena  & 14       & 24      & 210 & 11,926   \\ \midrule
    Total & 45       & 60      & null  & null   \\ \bottomrule
    \end{tabular}
\end{table}

\subsubsection{Real-World Attacked Projects}
The dataset includes 31 projects that have experienced real-world attacks, sourced from two reliable datasets:
\begin{itemize}
\item \textbf{11 Projects from GPTScan's DeFiHack Dataset~\cite{sun2023gptscan}:} This dataset, designed by GPTScan, originally contained 13 projects. We excluded two projects that were unrelated to oracle manipulation to ensure relevance.

\item \textbf{20 Projects from SOK~\cite{zhou2023sok}:} These projects, categorized in the original SOK paper as on-chain oracle manipulation, liquidity borrowing and depositing issues, and slippage exploitation observed between 2021 and 2022, are included to increase the dataset size.
\end{itemize}

\subsubsection{Code4Rena Projects}
Code4rena~\cite{code4rena} is a leading audit contest platform for pre-deployment projects. The platform engages project developers to commit bounties up to \$1M as incentives to draw participants from all over the world. 
Community experts selected and developers collaboratively review the submitted bug reports and reward the participants based on the severity and frequency of a particular bug submission. This incentive-driven process guarantees the integrity and credibility of the bugs reported, forming the ground truth for our study.

The dataset incorporates 14 projects from the Code4Rena platform:
\begin{itemize}
\item \textbf{6 Projects from Zhang et al.~\cite{zhang2023demystifying}:} These are derived from Zhang et al.'s original set of 11 price oracle manipulation projects on Code4Rena. To ensure fair comparison with GPTScan, we excluded two misclassified projects and three incomplete ones.
\item \textbf{8 Projects Newly Curated:} To enhance the robustness of the evaluation, we extended the dataset by including additional projects from the Code4Rena platform\footnote{\url{https://github.com/code-423n4/code423n4.com/tree/main/\_data/reports}} directly. 
We filtered projects from 2022 and 2023 using keywords such as \textit{frontrunning, slippage, flash loan, sandwich, price oracle}, and \textit{manipulation} in the audit reports. 
Each selected bug was manually checked to ensure it was a POM bug, and all projects relevant were confirmed to compile successfully. 
As a result, 8 projects were added to the dataset.
\end{itemize}
By focusing solely on price oracle manipulation, we enhanced both real-world attacked and Code4Rena datasets to ensure reliable evaluation. 

\subsection{Baseline} \label{baseline}
Since very few works have addressed the problem of price oracle manipulation, most existing analyses are post-mortem and rely on transaction data, which differs from our approach. Our goal is to prevent attacks before they occur.
To evaluate our framework, we select one state-of-the-art (SOTA) LLM-based tool, GPTScan, as our primary baseline. GPTScan utilizes an LLM (initially ChatGPT-3.5) to analyze pre-tagged functions potentially susceptible to specific vulnerability types through predefined scenarios and rules. The tool then applies static analysis to validate the LLM-generated findings and filter out false positives, thereby enhancing detection precision.
Given the deprecation of ChatGPT-3.5, we substituted it with ChatGPT-4o-mini, which OpenAI recommends as a more cost-effective and improved alternative.\footnote{https://openai.com/index/gpt-4o-mini-advancing-cost-efficient-intelligence/}
Additionally, the original GPTScan implementation suffered from unstructured LLM output, which hindered systematic analysis by the static analysis module. To address this limitation, we modified the tool's code to enforce structured JSON output supported by ChatGPT-4o-mini. These modifications ensure more consistent and reliable vulnerability assessments, improving the overall performance of the analysis pipeline.

To complement GPTScan, we design a zero-shot CoT prompt based on the assumption that a common LLM user, equipped with basic knowledge of prompt engineering and price oracle issues, could generate.
This baseline zero-shot CoT prompt generally follows the zero-shot prompting guideline outlined in the \textit{Prompt Engineering Guide}. \footnote{https://www.promptingguide.ai/}
We adopt the role where the model acts as \textit{an experienced expert on auditing price oracle manipulation problems}, with the task of identifying all potential vulnerabilities in a provided Solidity file.
The final prompt, structured for clarity and consistency, is shown below:

\begin{tcolorbox}[colback=gray!20, colframe=gray!50, title={Zero-shot CoT Prompt}, label=simplebaseline]
\textbf{System:} You are an experienced expert on auditing price oracle manipulation problems. Your task is to conduct a thorough audit on the provided solidity file step by step to identify all potential price oracle manipulation vulnerabilities. \\
\\
\textbf{User:} Analyze the smart contract delimited with \texttt{```}. Respond with all the vulnerabilities with the following JSON format: \{output format\}.
\end{tcolorbox}

\subsection{Model Choices and Hyperparameters} \label{subsec:settings}

\begin{table}
    \caption{Model Descriptions}
    \label{table:models}
    \begin{tabular}{@{}lcccc@{}}
    \toprule
    Models & Versions         & Knowledge Cutoff & Context Window & Max Tokens \\ \midrule
    ChatGPT-4o & 2024-05-13 & Oct 2023         & 128k         & 4,096       \\
    Claude3.5-Sonnet & 2024-06-20 & July 2024        & 200k         & 8,192         \\
    ChatGPT-4o-mini  &  2024-07-18    &  Oct 2023         & 128k       & 16,384      \\ 
    Claude3-haiku  &  2024-07-18    & Aug 2023        & 200k       & 	4,096       \\ \bottomrule
    \end{tabular}%
\end{table}

We utilized four language models, as described in Table~\ref{table:models}. These models were selected to represent the latest advancements in industry that are both accessible and affordable for everyday users.
The \textit{Knowledge Cutoff} column is the cutoff date of the training data, indicating the recency of the knowledge embedded in the model.
The \textit{Context Window} is the maximum combined length of input tokens and output tokens that the model can process in a single query.
The \textit{max\_tokens} is the upper limit of tokens that the model can generate as output. 
Although these models allow for higher token generation, the maximum output token limit was set to $1,024$ in our experiments to prioritize longer input.

\begin{table}[]
\caption{Model Parameters}
\label{table:modelparams}
\begin{tabular}{@{}lccc@{}}
\toprule
Parameters  & Knowledge Synthesizer & Generator & Auditor \\ \midrule
temperature & 0                     & 0         & 1.0     \\
top\_p      & 1.0                   & 1.0       & 1.0     \\ \bottomrule
\end{tabular}
\end{table}

In this study, we evaluated the performance of the models by adjusting two key parameters: \textit{temperature} and \textit{top\_p}. The \textit{temperature} parameter controls the randomness of the output, with higher values promoting greater variability and lower values yielding more deterministic results. In contrast, \textit{top\_p} applies nucleus sampling, where only tokens contributing to the top \textit{p} probability mass are considered. For instance, a \textit{top\_p} value of 0.1 restricts the model to tokens comprising the top 10\% of cumulative probability. OpenAI generally advises modifying either \textit{top\_p} or \textit{temperature}, but not both simultaneously.

In this evaluation, we set the \textit{top\_p} parameter to its default value of 1.0, ensuring that the full probability mass was considered. The \textit{temperature} parameter was configured as follows: a value of 0 was applied for the \textit{Knowledge Synthesizer} and \textbf{Prompt Generator} to prioritize accuracy and consistency, while the \textbf{Auditor} was assigned a value of 1.0 to encourage diverse and comprehensive vulnerability identification, thereby fully leveraging the model’s capabilities. All other parameter settings adhered to their default values as outlined in the OpenAI API documentation~\cite{openaiparams}.

To mitigate the impact of randomness inherent in the models, each prompt was executed on the dataset three times. The models were instructed to identify vulnerable functions and provide detailed explanations of the vulnerabilities. The outputs were formatted in JSON to facilitate efficient post-processing.

\subsection{Identifying the Best Configuration for \tool} ~\label{subsec:baselines}
This section is to establish the most effective combination of components for the POM detection task.
We start by evaluating the performance of the \textit{Auditor} component in isolation. 
Subsequently, we enhance the setup by incorporating the \textit{Prompt Generator} with human-curated knowledge to improve performance.
Finally, we automate the knowledge extraction process by integrating the \textit{Knowledge Synthesizer} for better usability.

\subsubsection{Identifying the Best Auditor Model}

We assessed the zero-shot CoT prompt (described in Section~\ref{simplebaseline}) on the DeFiHacks dataset to identify the most suitable model as the auditor for detecting POM issues. For comparison, we also evaluated GPTScan on the same dataset.
The DeFiHacks dataset was chosen because it allows for more efficient manual verification of the outputs, facilitating the identification of optimal parameter combinations. 

The results are summarized in Table~\ref{table:baselines}.  The metrics used to assess performance include \textbf{False Negatives (FN)}, which stands for the number of vulnerabilities incorrectly classified as safe; \textbf{True Positives (TP)}, the number of the vulnerabilities correctly identified and \textbf{False Positives (FP)}, the number of instances incorrectly flagged as vulnerable but are actually safe.
We evaluated performance using precision, recall, and F1 score as key indicators. \textbf{Precision} measures the accuracy of positive predictions, while \textbf{Recall} assesses the model's ability to identify all relevant instances.
The \textbf{F1 score}, as the harmonic mean of precision and recall, offers a balanced metric, particularly valuable for imbalanced class distributions. For detailed formulas, refer to Appendix~\ref{app:formulas}.


\begin{table}
\caption{Average Performance of Baselines on DeFiHacks}
\label{table:baselines}
\begin{tabular}{@{}l|c|cccc@{}}
\toprule
          & GPTScan & \multicolumn{4}{c}{Zero-shot CoT}                                      \\
Models    & 4o-mini & 4o    & Sonnet & 4o-mini                                         & Haiku \\ \midrule
FN        & 26.67   & 25.67 & 29.67  & 14.33                                           & 26.33 \\
TP        & 9.33    & 10.33 & 6.33   & 21.67                                           & 9.67  \\
FP        & 20.67   & 19.67 & 23.00  & 58.33                                           & 19.33 \\ \midrule
Recall    & 0.259   & 0.287 & 0.176  & 0.602                                           & 0.269 \\
Precision & 0.311   & 0.344 & 0.216  & 0.271                                           & 0.333 \\
F1        & 0.283   & 0.313 & 0.194  & \color[HTML]{C834FC} 0.374 & 0.297 \\ \bottomrule
\end{tabular}
\end{table}

As shown in Table~\ref{table:baselines}, almost all the models achieved higher F1 Score performance compared to GPTScan. Among the evaluated models, 4o-mini from the zero-shot CoT approach demonstrated superior performance, achieving the highest F1 Score (0.374) on the DeFiHacks dataset, outperforming GPTScan (0.283) and other models, including 4o version. Similarly, Claude's Haiku outperformed its more advanced counterpart, Sonnet, in terms of F1 Score (0.297 vs. 0.194).
A detailed review of the results shows that more complex and advanced models tend to generate more conservative output, while this leads to lower false positive, it produces fewer overall predictions,  which is more likely to miss true vulnerabilities.

Notably, while GPTScan significantly reduced false positives, it also increased false negatives. This trade-off resulted in a marginally higher precision (0.311 vs. 0.271) but at the expense of a markedly lower recall (0.259 vs. 0.602).


\begin{tcolorbox}[colback=gray!20!white, colframe=gray!75!black, boxsep=5pt, arc=4pt, boxrule=1pt, left=0pt, right=0pt]
\textbf{Finding 1:}
4o-mini demonstrates potential as a better auditor. "Mini" versions like 4o-mini and Claude's Haiku outperformed their flagship counterparts, with 4o-mini achieving the highest F1 Score (0.374) among all models. 
\end{tcolorbox}

\begin{tcolorbox}[colback=gray!20!white, colframe=gray!75!black, boxsep=5pt, arc=4pt, boxrule=1pt, left=0pt, right=0pt]
In response to \textbf{RQ1}: Despite advancements in model performance, both GPTScan and the zero-shot CoT approaches show limited effectiveness in reliably detecting price oracle problems, emphasizing the need for further refinement.
\end{tcolorbox}

To clarify the discrepancy between our evaluation and the report from GPTScan's paper, it is important to note that the metrics used in both evaluations differ. In GPTScan's report, they emphasize the identification of vulnerability types across projects.
For instance, in the project \textit{Hack-20210603-PancakeHunny}, 7 vulnerabilities of \textit{Flash Loan Price (FLP)} were found in their report, but only 1 true positive (TP) was counted in their precision and recall calculations. This difference in how vulnerabilities are calculated and counted contributes to the significant variation in the results presented here.

\subsubsection{Identifying the Best Prompt Generator with Manually Curated Knowledge}

Building on the results of the previous section, where "4o-mini" was identified as the best auditor model using the zero-shot CoT prompt, we now explore varying models for the prompt generator to determine the optimal combination for detecting POM vulnerabilities. This section also tries to validate the performance of the best auditor model identified again.

To balance computational efficiency with reliability, we adopted a strategic optimization approach.
Specifically, we stabilized the domain knowledge component by integrating human-curated expertise into the prompt generator. This refinement allowed us to constrain the experimental space and focus on identifying optimal combinations of models for vulnerability detection.
With the optimal model combinations identified, we can systematically reduce manual intervention with the knowledge synthesizer and progressively automate our vulnerability detection framework, ultimately advancing towards a more autonomous and robust oracle manipulation detection system.
The curated domain knowledge that underpins this optimization is presented in Appendix~\ref{app:humanknowledge}.

\begin{figure}[h]
  \centering
  \includegraphics[width=\linewidth]{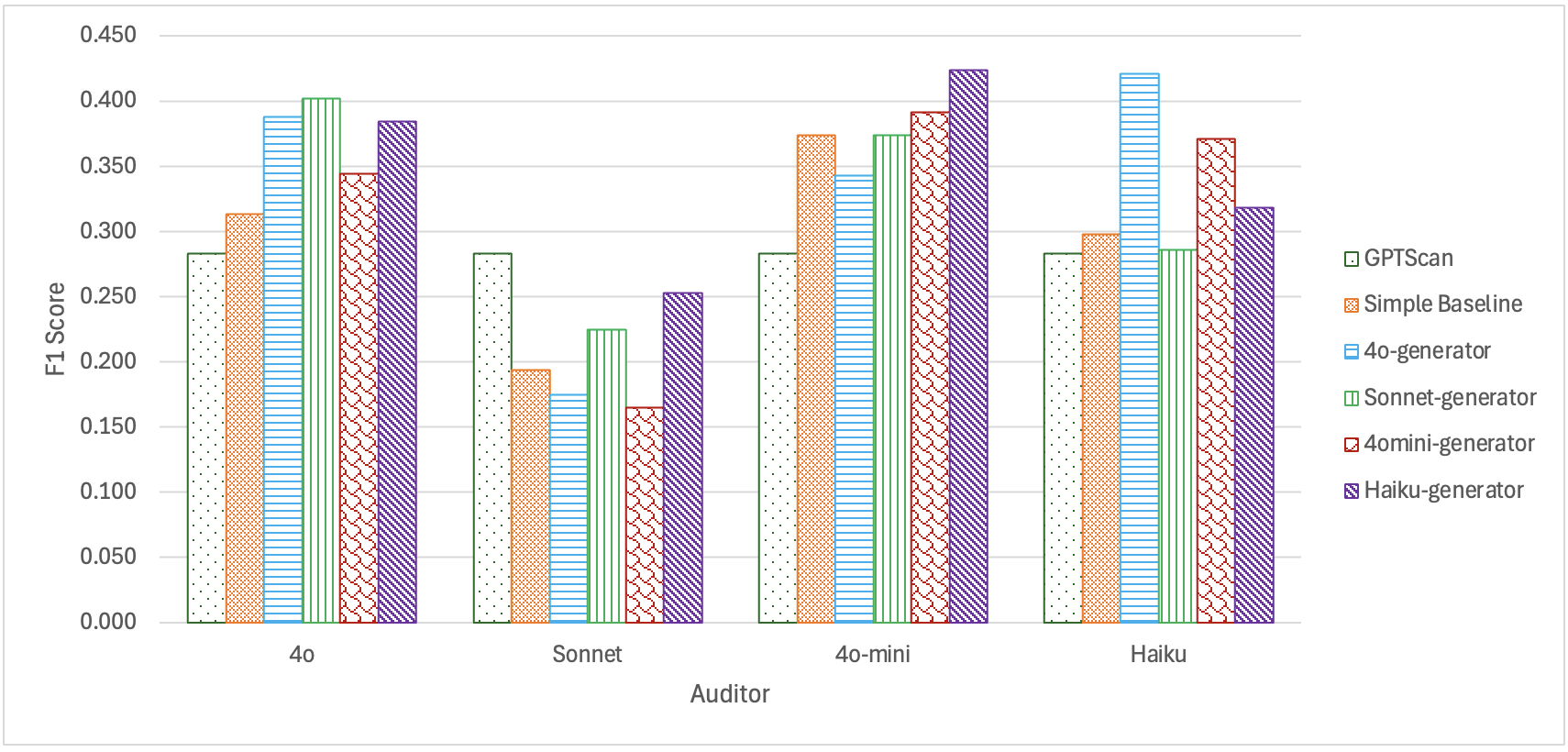}
  \caption{F1 Score of Varying Prompt Generator and Auditor with Human-curated Knowledge.}
  \Description{Performance of varying models for prompt generator and auditor.}
  \label{fig:varymodel}
\end{figure}

The results are presented in Figure~\ref{fig:varymodel}, which illustrates a comparative analysis of different model configurations applied to the DeFiHacks dataset. The raw data for these results is provided in Table~\ref{table:aigenprompt}.
In this figure, the y-axis represents performance, measured by the F1-score. 
Each bar cluster in the x-axis denote the choice of auditor model (ranges in "4o", "Sonnet", "4o-mini" and "Haiku").
Within each cluster, different bars represent different choices of the paired generator models (ranges in "4o-generator", "Sonnet-generator", "4o-mini-generator" and "Haiku-generator") as well as the two baselines (GPTScan and zero-shot CoT baseline).
Note that, we stick to use 4o-mini for GPTScan in all the comparisons, and the results reported in Section~\ref{subsec:baselines} is reused for the zero-shot CoT.

\begin{table}
\caption{Average Performance of Varying Prompt Generator and Auditor with Human-curated Knowledge}
\label{table:aigenprompt}
\begin{adjustbox}{width=\textwidth}
\begin{tabular}{@{}l|cccc|cccc|cccc|cccc@{}}
\toprule
Prompt Gen & 4o    & Sonnet & 4o-mini & Haiku & 4o     & Sonnet & 4o-mini & Haiku  & 4o      & Sonnet  & 4o-mini & Haiku   & 4o    & Sonnet & 4o-mini & Haiku \\ 
Auditor          & 4o    & 4o     & 4o      & 4o    & Sonnet & Sonnet & Sonnet  & Sonnet & 4o-mini & 4o-mini & 4o-mini & 4o-mini & Haiku & Haiku  & Haiku   & Haiku \\ \midrule
FN               & 23.33 & 23.00  & 25.33   & 24.67 & 30.33  & 28.67  & 30.67   & 27.67  & 18.00   & 12.33   & 12.67   & 13.20   & 22.33 & 26.67  & 24.00   & 25.67 \\
TP               & 12.67 & 13.00  & 10.67   & 11.33 & 5.67   & 7.33   & 5.33    & 8.33   & 18.00   & 23.67   & 23.33   & 22.80   & 13.67 & 9.33   & 12.00   & 10.33 \\
FP               & 16.67 & 15.67  & 15.33   & 11.67 & 23.33  & 22.00  & 23.33   & 21.67  & 51.00   & 67.00   & 60.00   & 48.80   & 15.33 & 20.00  & 16.67   & 18.67 \\ \midrule
Recall           & 0.352 & 0.361  & 0.296   & 0.315 & 0.157  & 0.204  & 0.148   & 0.231  & 0.500   & 0.657   & 0.648   & 0.633   & 0.380 & 0.259  & 0.333   & 0.287 \\
Precision        & 0.432 & 0.453  & 0.410   & 0.493 & 0.195  & 0.250  & 0.186   & 0.278  & 0.261   & 0.261   & 0.280   & 0.318   & 0.471 & 0.318  & 0.419   & 0.356 \\
F1               & 0.388 & 0.402  & 0.344   & 0.384 & 0.174  & 0.224  & 0.165   & 0.253  & 0.343   & 0.374   & 0.391   & {\color[HTML]{C834FC} 0.424}   & {\color[HTML]{C834FC} 0.421} & 0.286  & 0.371   & 0.318 \\ \bottomrule
\end{tabular}
\end{adjustbox}
\end{table}

As demonstrated in the Figure, the performance varies significantly across different combinations of prompt generators and auditor models.
Nevertheless, our framework consistently outperforms the zero-shot CoT across 12 out of 16 model configurations, underscoring the efficacy of leveraging human-curated knowledge and optimized, context-aware prompts. 
Among the results obtained and showed in Table~\ref{table:aigenprompt} and Figure~\ref{fig:varymodel}, two combinations stand out: 1). "Haiku" as the prompt generator paired with "4o-mini" as the auditor: This combination achieved the highest F1 score of 0.424, representing a 13.4\% improvement over the zero-shot CoT of 4o-mini. It also attained the highest recall (0.633), making it ideal for identifying a broader range of vulnerabilities.
2). "4o" as the prompt generator paired with "Haiku" as the auditor: This combination achieved an F1 score of 0.421, a 41.8\% improvement over the zero-shot CoT of Haiku. It exhibited higher precision, making it preferable for scenarios where minimizing false positives is critical.
These results highlight the synergistic benefits of pairing complementary model architectures, enabling tailored optimization for varying detection priorities.

\begin{tcolorbox}[colback=gray!20!white, colframe=gray!75!black, boxsep=5pt, arc=4pt, boxrule=1pt, left=0pt, right=0pt]
\textbf{Finding 2:} 
Our framework, enhanced with human-curated knowledge, improves upon the baseline prompt in most cases (12 out of 16). Notably, two combinations—‘Haiku-generator’ with the ‘4o-mini’ auditor and ‘4o-generator’ with the ‘Haiku’ auditor—achieved the highest F1-scores of 0.424 and 0.421, compared to the baseline's best of 0.374.
\end{tcolorbox}

\begin{tcolorbox}[colback=gray!20!white, colframe=gray!75!black, boxsep=5pt, arc=4pt, boxrule=1pt, left=0pt, right=0pt]
In response to \textbf{RQ2}: \tool~identifies two combinations that outperform the best baseline performance with human-curated knowledge, demonstrating the efficacy of \tool.
\end{tcolorbox}

\subsubsection{Identifying the Best Knowledge Synthesizer for Automation}
With the best auditor and prompt generator combinations identified, the next step is to optimize the knowledge synthesizer for summarizing domain-specific information, ultimately aiming for a fully automated vulnerability detection framework. The knowledge synthesizer plays a critical role in utilizing domain knowledge from external domain-specific sources, and integrating them into the detection pipeline.
To systematically evaluate the knowledge synthesizer's effectiveness, we fixed the two top-performing combinations of Prompt Generator and Auditor from \textbf{Finding 2} (Haiku-generator with 4o-mini auditor, and 4o-generator with Haiku auditor) while varying the knowledge synthesizer models. Performance metrics included consistency, accuracy, and the overall F1-score.

As summarized in Table~\ref{table:aigenknowledge}, the combination of Haiku as the Knowledge Synthesizer, Prompt Generator, and Auditor (Haiku-Haiku-4o-mini) achieved the best performance, with an F1-score of 0.426. Remarkably, this result slightly surpasses that of the human-curated knowledge framework (0.426 vs. 0.424), demonstrating the potential of fully automated domain knowledge synthesis in enhancing vulnerability detection capabilities.
This improvement is particularly noteworthy as it highlights the ability of automated frameworks to match—and even exceed—manual approaches. Additionally, the Haiku-Haiku-4o-mini configuration achieved this result without requiring extensive manual intervention, marking a significant step toward a robust and autonomous system.

\begin{table}
\caption{Average Performance of Varying Knowledge Synthesizer}
\label{table:aigenknowledge}
\begin{tabular}{@{}l|cccc|cccc@{}}
\toprule
Knowledge Synthesizer & 4o      & Sonnet  & 4o-mini & Haiku   & 4o    & Sonnet & 4o-mini & Haiku \\ 
\midrule
Prompt Generator      & Haiku   & Haiku   & Haiku   & Haiku   & 4o    & 4o     & 4o      & 4o    \\
Auditor               & 4o-mini & 4o-mini & 4o-mini & 4o-mini & Haiku & Haiku  & Haiku   & Haiku \\
\midrule
FN                    & 11.00   & 10.00   & 14.00   & 12.00   & 26.00 & 23.00  & 26.00   & 25.00 \\
TP                    & 25.00   & 26.00   & 22.00   & 24.00   & 10.00 & 13.00  & 10.00   & 11.00 \\
FP                    & 86.00   & 74.67   & 56.00   & 52.67   & 18.00 & 17.00  & 20.00   & 19.00 \\ 
\midrule
Recall                & 0.694   & 0.722   & 0.611   & 0.667   & 0.278 & 0.361  & 0.278   & 0.306 \\
Precision             & 0.225   & 0.258   & 0.282   & 0.313   & 0.357 & 0.433  & 0.333   & 0.367 \\
F1                    & 0.340   & 0.380   & 0.386   & {\color[HTML]{C834FC} 0.426}   & 0.313 & 0.394  & 0.303   & 0.333 \\ \bottomrule
\end{tabular}
\end{table}

\begin{tcolorbox}[colback=gray!20!white, colframe=gray!75!black, boxsep=5pt, arc=4pt, boxrule=1pt, left=0pt, right=0pt] 
\textbf{Finding 3:}
The combination of Haiku-Haiku-4o-mini achieved the highest F1 score (0.426), slightly outperforming the human-curated knowledge framework (0.424). This result underscores the potential of fully automated domain knowledge synthesis for advancing vulnerability detection.
\end{tcolorbox}

\begin{tcolorbox}[colback=gray!20!white, colframe=gray!75!black, boxsep=5pt, arc=4pt, boxrule=1pt, left=0pt, right=0pt] 
\textbf{Finding 4:}
The best performance is achieved using a combination of less complex models (Haiku-Haiku-4o-mini), demonstrating that larger models do not necessarily lead to better results. 
This finding highlights the potential for accessibility and suggests the feasibility of using alternative models similar in size and complexity.
\end{tcolorbox}

\begin{tcolorbox}[colback=gray!20!white, colframe=gray!75!black, boxsep=5pt, arc=4pt, boxrule=1pt, left=0pt, right=0pt]
In response to \textbf{RQ3}:  the \textit{Knowledge Synthesizer} automates knowledge extraction and slightly surpasses human-curated knowledge in both recall and precision, enhancing the framework's performance, while reducing manual effort.
\end{tcolorbox}

\subsection{Comparison of \tool~and GPTScan on Code4Rena}
To evaluate the effectiveness of \tool~on fully developed projects with all supporting files, we applied the best-performing configuration identified in previous sections—Haiku as the Knowledge Synthesizer and Prompt Generator, paired with 4o-mini as the Auditor—on the Code4Rena dataset. 
This dataset presents additional challenges due to its complexity and diverse set of components, providing a rigorous test of the system's capabilities. 
For comparison, we also evaluated the state-of-the-art tool GPTScan on the same dataset.

\begin{table}[]
\caption{Performance of \tool~and GPTScan on Code4rena Dataset}
\label{table:comparisonc4}
\begin{tabular}{@{}lccc|ccc@{}}
\toprule
Metrics    & FN & TP & FP    & Recall & Precision & F1    \\ \midrule
\tool      & 11 & 13 & 128.3 & 0.54   & 0.092     & 0.157 \\
GPTScan    & 21 & 3  & 29    & 0.13   & 0.094     & 0.107 \\ \bottomrule
\end{tabular}
\end{table}

The performance comparison between \tool~ and GPTScan are summarized in Table~\ref{table:comparisonc4}.
\tool~demonstrated a significant improvement in recall (0.54) compared to GPTScan (0.13), indicating a superior ability to detect vulnerabilities in the dataset. 
This higher recall highlights the effectiveness of the LLM generated knowledge synthesis and optimized prompt generation.
However, \tool's precision (0.092) remains comparable to GPTScan (0.094), reflecting the challenges in reducing false positives in complex datasets.
Despite the trade-offs, \tool~achieved a higher overall F1-score of 0.157 compared to GPTScan's 0.107, underscoring its improvements on complex dataset.

These findings highlight that a simple and general framework \tool~ can outperform state-of-the-art tools like GPTScan, even in complex and challenging datasets. 
This positions \tool~as a promising solution for advancing vulnerability detection frameworks.

\begin{tcolorbox}[colback=gray!20!white, colframe=gray!75!black, boxsep=5pt, arc=4pt, boxrule=1pt, left=0pt, right=0pt] 
\textbf{Finding 5:}
\tool~outperforms GPTScan on the challenging Code4Rena dataset, achieving higher recall (0.54 vs. 0.13) and F1-score (0.157 vs. 0.107). However, its precision remains on par with GPTScan, reflecting the need for further refinements to reduce false positives.
\end{tcolorbox}

%% file: chapters/5_related.tex
\section{Related Work}\label{sec:related}

Smart contract vulnerability detection has advanced significantly, with numerous tools and techniques proposed to address security issues. Despite these efforts, detecting and mitigating manipulative behaviors in price oracles remains a persistent challenge. Existing approaches to this problem can be broadly classified into static analysis, dynamic analysis, machine learning-based methods, and emerging techniques leveraging large language models (LLMs).

\subsection{Static Analysis} \label{Static Analysis}
Static analysis techniques examine the source code or bytecode of smart contracts without execution, employing methods such as symbolic execution, formal verification, and pattern matching to identify vulnerabilities. Several tools have been developed for various types of vulnerabilities. For instance, Oyente\cite{luu2016oyente} uses symbolic execution to detect issues such as reentrancy, transaction order dependency, suicidal contracts, and integer overflows. SmartCheck\cite{tikhomirov2018smartcheck} applies rule-based techniques to identify vulnerabilities and bad practices in Solidity contracts, while Slither~\cite{feist2019slither} combines dataflow analysis, taint analysis, and pattern matching to detect a wide range of vulnerabilities efficiently.
Formal verification methods further enhance static analysis by modeling smart contract behavior using formal languages and verifying properties with SMT solvers or theorem provers. Examples include VeriSmart\cite{so2020verismart} and sVerify\cite{gao2021sverify}, which are tailored for smart contract verification against predefined specifications.

However, few works focus explicitly on price oracle manipulation vulnerabilities. Recent research has started addressing this gap:
Foray~\cite{wen2024foray} is an attack synthesis framework for DeFi protocols that uses a domain-specific language to convert smart contracts into token flow graphs. While it identifies strategic paths and synthesizes attacks via symbolic compilation, its focus is limited to four specific types of logical flaws, which partially overlap but do not fully align with our target vulnerabilities.
OVer~\cite{deng2024safeguarding} employs symbolic analysis to model DeFi protocol behavior under skewed oracle inputs, identifying secure parameters and generating guard statements to mitigate manipulation attacks. While effective, its focus on optimizing parameters for specific protocols limits its generalizability to broader applications.
VeriOracle~\cite{mo2023toward} introduces a formal verification framework that deploys a semantic model on the blockchain to monitor smart contract states and detect problematic price feed transactions in real time.
DeFiTainter~\cite{kong2023defitainter} leverages innovative mechanisms to construct call graphs and semantically track inter-contract taint data for detecting price manipulation vulnerabilities. 
However, they both require extensive on-chain transaction data, demanding significant resources and differing from our approach.

\subsection{Dynamic Analysis} \label{Dynamic Analysis}
Dynamic analysis techniques execute smart contract code and monitor its runtime behavior to identify vulnerabilities. These methods often employ fuzzing, symbolic execution, and runtime monitoring to detect issues such as assertion failures, overflows, and frozen ether. Tools like Mythril\cite{mueller2024mythril}, Manticore\cite{mossberg2019manticore}, sFuzz\cite{nguyen2020sfuzz}, and ContractFuzzer\cite{jiang2018contractfuzzer} have been widely used for identifying common vulnerabilities.

Despite their success, traditional dynamic analysis tools have rarely addressed price oracle manipulation vulnerabilities. Only a few works have specifically targeted this challenge.
DeFiRanger~\cite{wu2021defiranger} recovers high-level DeFi semantics from raw Ethereum transactions and identifies price oracle manipulation attacks through pattern matching. However, its approach is post-mortem, as it can only detect observed attack transactions, limiting its usefulness for proactive vulnerability detection.
ProMutator~\cite{wang2021promutator} models typical DeFi usage patterns by analyzing existing transactions and simulates potential price manipulation attacks through mutated transactions. This approach effectively identifies weak points in oracle systems before exploitation. However, accurately modeling DeFi transaction patterns is challenging, especially for novel attack vectors, and its simulation-based method requires significant computational resources, impacting scalability and real-time applicability.
DeFiPoser~\cite{zhou2021defiposer} employs a dual approach: DEFIPOSER-ARB for identifying arbitrage opportunities and SMT solvers to create logical models for detecting complex profitable transactions. While it can uncover new vulnerabilities in real time, the system relies on manual and costly modeling of DeFi protocols, making it resource-intensive. Furthermore, its effectiveness may be limited by the rapid evolution of DeFi protocols, requiring frequent updates to maintain accuracy and relevance.

These limitations highlight the advantages of our AI-driven framework, which eliminates reliance on expert knowledge and enhances scalability, effectively 
overcoming the inefficiencies and adaptability challenges in existing 
methods.

\subsection{Machine Learning-based Methods} \label{Machine Learning-based Methods}
Machine learning-based methods have gained traction in recent years for smart contract vulnerability detection. These approaches typically involve extracting features from the contract's source code or bytecode and training classifiers or deep learning models to predict the presence of vulnerabilities. Some notable works in this domain include ContractWard~\cite{wang2020contractward}, which trains a classifier based on features extracted from the contract's bytecode, and the hybrid approach proposed by Liu et al.~\cite{liu2021combining}, which combines pure neural networks with interpretable graph features and expert patterns.
Graph neural networks have also been explored for smart contract vulnerability detection. These approaches represent the contract's control flow graph or data dependency graph as a graph-structured data and apply graph neural networks to learn vulnerability patterns. EtherGIS~\cite{zeng2022ethergis} is an example of a vulnerability detection framework that utilizes graph learning features to detect vulnerabilities in Ethereum smart contracts.

While these methods achieve high accuracy for various vulnerabilities, they rely heavily on large labeled datasets and often struggle with novel or unseen patterns. Moreover, limited work has specifically addressed price oracle manipulation vulnerabilities, leaving a gap that requires innovative solutions.

\subsection{Large Language Model-based Methods} \label{Large Language Model-based Methods}
Recent advancements in large language models (LLMs) have opened up new possibilities for smart contract vulnerability detection. LLMs, such as GPT, have demonstrated remarkable capabilities in understanding and generating human-like text, and researchers have begun exploring their application in smart contract analysis.
Gao et al.\cite{gao2024unveiling} explored LLMs for detecting complex bugs, including price oracle manipulation, using diverse prompts. However, this early work demonstrated limited performance, making it unsuitable for practical use. Similarly, Issac et al.\cite{david2023you} evaluated ChatGPT-4 and Claude for smart contract audits, identifying logic flaws and coding errors but reporting an unacceptably high false positive rate (95\%), which hinders real-world adoption. GPTLens~\cite{hu2023large} proposed an adversarial framework leveraging LLMs in dual roles to enhance detection accuracy, but its effectiveness on price oracle manipulation vulnerabilities remains limited.
GPTScan~\cite{sun2023gptscan} combines GPT with program analysis techniques to identify logic vulnerabilities in smart contracts. By leveraging GPT's code understanding and static confirmation, GPTScan reduces false positives and achieves high precision and recall in terms of vulnerability type detection across diverse datasets. 

In contrast, our work develops a fully LLM-driven approach focused on prompt engineering for detecting POM vulnerabilities. 
By utilizing domain-specific knowledge extraction and context-aware prompt generation, we enable LLMs to automatically identify manipulation patterns. Our method is user-friendly, generalizable, and provides actionable feedback by leveraging LLMs' capacity to understand the contextual 
nuances of price oracle manipulations.

%% file: chapters/6_conclusion.tex
\section{Conclusion and Future Work}\label{sec:conclusion}
POM attacks represent a pressing challenge in the DeFi space. \tool~provides an innovative, LLM-driven solution that automates the detection of these vulnerabilities by leveraging domain-specific knowledge and context-aware prompt generation. By streamlining the analysis process and ensuring actionable outputs, \tool~offers a scalable and effective approach to safeguarding DeFi ecosystems.

For future work, several promising directions can be explored to enhance and expand the capabilities of \tool:

\begin{itemize}
\item Extension to Additional Problems: \tool's framework can be expanded to address other DeFi issues, such as the accounting problem, privilege escalation, and inconsistent state updates.

\item Automating Knowledge Synthesis: To reduce manual effort, future versions of \tool~will enhance automation in knowledge synthesis, potentially using Retrieval-augmented Generation (RAG)~\cite{gao2023retrieval} for efficient data extraction and summarization.

\item Instruction-tuned LLM: This effort may include the construction of suitable instruction-following dataset and further supervised fine-tuning of the prompt generator LLM and/or the auditor LLM~\cite{longpre2023flan} to better follow the instructions and prompts given and align the model's behave towards desired output.

\item Reducing False Alarms: Enhancing usability by exploring advanced and fine-grained prompt techniques, such as adaptive prompts and context-aware filtering, to minimize false alarms while maintaining high detection accuracy.
\end{itemize}

By pursuing these directions, \tool~can continue to evolve, offering comprehensive protection against a growing array of vulnerabilities in the DeFi ecosystem, while enhancing its efficiency and user-friendliness.

%% file: chapters/appendix.tex
\appendix

\section{Appendix: Prompts of \tool} \label{app:componentprompts}
\subsection{Prompt Generated by Prompt Generator}
\begin{tcolorbox}[colback=gray!20, colframe=gray!50, title={Prompt Generated by Prompt Generator}, label=cotprompt]
description: Detect price oracle manipulation vulnerabilities in a smart contract,  steps:
step: 1, description: Identify the price oracle used by the smart contract. Determine if the price oracle is a decentralized exchange (DEX) or another external price feed.,      
    questions: What is the price oracle used by the smart contract?       
    Is the price oracle a decentralized exchange (DEX) or another external price feed?
step: 2, description: Analyze the smart contract's reliance on the price oracle. Determine if the contract uses the price provided by the oracle directly, without any additional validation or cross-checking.,     
    questions: Does the smart contract rely solely on the price provided by the price oracle, without any additional validation?
    Are there any mechanisms in place to detect and mitigate price manipulation in the price oracle?  
step: 3, description: Evaluate the potential impact of a price oracle manipulation attack on the smart contract. Determine if the attack could lead to financial losses or unfair advantages for the attacker.,      
    questions: What are the potential financial consequences if the price oracle is manipulated?       
    Could the price oracle manipulation lead to unfair advantages for the attacker? 
step: 4, description: Provide recommendations to mitigate the price oracle manipulation vulnerability, such as using multiple price oracles, implementing price validation mechanisms, or using more robust price feeds.,      
    questions: What are the recommended mitigation strategies to address the price oracle manipulation vulnerability?
    How can the smart contract be modified to reduce the risk of price oracle manipulation attacks?
\end{tcolorbox}

\section{Appendix: Knowledge}

\subsection{Human-curated Domain Knowledge} \label{app:humanknowledge}
\begin{tcolorbox}[colback=gray!20!white, colframe=gray!75!black, boxsep=5pt, arc=4pt, boxrule=1pt, left=0pt, right=0pt, title={Human-curated Domain Knowledge}, label=humanknowledge]
Liquidity pools provide the underlying liquidity for decentralized exchanges (DEXs) like Uniswap by holding token reserves. These pools can also act as on-chain price oracles, offering real-time price data to other applications. When discussing smart contracts vulnerable to price oracle manipulation, we focus on two main categories of vulnerabilities:

1. Vulnerabilities in DEXes and Liquidity Pools/reserves: a) Lack of Slippage Tolerance: Users may suffer unexpected losses due to significant price fluctuations during swaps or trades if the smart contracts do not have adequate slippage tolerance settings. b) Susceptibility to Front-Running or Sandwich Attacks: Smart contracts that do not mitigate front-running or sandwich attacks can expose users to losses when malicious actors manipulate transaction order and timing.

2. Vulnerabilities in Price-Dependent Applications: a) Price Oracle Manipulation: Smart contracts that rely on external price oracles can be manipulated through tampering with the price feed, leading to incorrect asset valuations and financial loss. b) Unfair asset valuation: Poorly designed smart contracts may allow malicious users to manipulate asset valuations, even if the oracle price is accurate. Inadequate safeguards can enable unfair trading practices, disadvantaging other users.
\end{tcolorbox}

\section{Appendix: Illustration of POM} \label{app:oraclemanip}
As discussed in Section~\ref{sec:introduction}, price oracles are integral to the functionality of DeFi applications but are also vulnerable to manipulation. One example of POM is depicted in Figure~\ref{fig:priceoracle}, which is excerpted from project \textit{Behodler}\footnote{\url{https://code4rena.com/reports/2022-01-behodler}}.

\begin{figure}[h]
\centering
\begin{lstlisting}[language=Solidity]
function burnAsset(address asset, uint256 amount) public isLive incrementFate {
    require(assetApproved[asset], "LimboDAO: illegal asset");
    address sender = _msgSender();
    require(ERC677(asset).transferFrom(sender, address(this), amount), "LimboDAO: transferFailed");
    uint256 fateCreated = fateState[_msgSender()].fateBalance;
    uint256 actualEyeBalance = IERC20(domainConfig.eye).balanceOf(asset);
    require(actualEyeBalance > 0, "LimboDAO: No EYE");
    uint256 totalSupply = IERC20(asset).totalSupply();
    uint256 eyePerUnit = (actualEyeBalance * ONE) / totalSupply;
    uint256 impliedEye = (eyePerUnit * amount) / ONE;
    fateCreated = impliedEye * 20;
    fateState[_msgSender()].fateBalance += fateCreated;
    emit assetBurnt(_msgSender(), asset, fateCreated);
}
\end{lstlisting}
\caption{Example function \texttt{burnAsset} with potential flash loan attack vulnerability.}
\label{fig:priceoracle}
\end{figure}

The \textbf{burnAsset} function is designed to remove tokens from circulation by burning assets and crediting \textit{Fate} tokens to users. \textit{Fate} tokens serve as a governance currency within the ecosystem, granting holders voting power. This function interacts with EYE-based asset tokens, but the asset pricing formula is vulnerable to flash loan manipulation.

Consider a scenario where there are 1000 EYE and 1000 LINK tokens in a UniswapV2 LINK-EYE pool. The pool’s total supply is 1000, and the attacker holds 100 LP tokens. If the attacker calls the \textbf{burnAsset} function to burn their 100 LP tokens, with the formula in line 9-11, he can earn $1000 \times 100/1000 \times 20 = 2000$ amount of \textit{Fate}.
Here, $1000$ is the \textbf{actualEyeBalance} and $1000$ is the pool's total LP supply. Thus, the attacker rightfully receives 2000 \textit{Fate} tokens.

However, the attacker can exploit the system by swapping in 1000 EYE and receiving 500 LINK from the pool (according to $x \times y = k$, ignoring fees for simplicity). The pool then contains 2000 EYE and 500 LINK tokens. The \textbf{actualEyeBalance} becomes 2000, while the pool's total LP supply and the attacker's LP tokens remain at 1000 and 100, respectively. After this manipulation, the attacker can call the \textbf{burnAsset} function to burn their LP tokens and receive $2000 \times 100/1000 \times 20 = 4000$ amount of \textit{Fate} tokens.Subsequently, the attacker can swap 500 LINK back into the pool to retrieve their 1000 EYE.
Ultimately, the attacker incurs only the transaction fee, yet they gain double the \textit{Fate} tokens (4000) compared to the legitimate amount (2000). With this increased \textit{Fate}, the attacker gains more voting power to influence the system’s decisions or can convert \textit{Fate} to other tokens for direct profit.

This example illustrates how the ratio of pool tokens can be manipulated through flash loans to exploit price oracles, leading to significant imbalances and vulnerabilities in DeFi applications.

\section{Appendix: Formulas} \label{app:formulas}
\[
\text{Precision} = \frac{\text{True Positives (TP)}}{\text{True Positives (TP)} + \text{False Positives (FP)}}
\]

\[
\text{Recall} = \frac{\text{True Positives (TP)}}{\text{True Positives (TP)} + \text{False Negatives (FN)}}
\]

\[
F_1\text{-score} = 2 \cdot \frac{\text{Precision} \cdot \text{Recall}}{\text{Precision} + \text{Recall}}
\]

%% file: main.bbl
\begin{thebibliography}{44}


\ifx \showCODEN    \undefined \def \showCODEN     #1{\unskip}     \fi
\ifx \showDOI      \undefined \def \showDOI       #1{#1}\fi
\ifx \showISBNx    \undefined \def \showISBNx     #1{\unskip}     \fi
\ifx \showISBNxiii \undefined \def \showISBNxiii  #1{\unskip}     \fi
\ifx \showISSN     \undefined \def \showISSN      #1{\unskip}     \fi
\ifx \showLCCN     \undefined \def \showLCCN      #1{\unskip}     \fi
\ifx \shownote     \undefined \def \shownote      #1{#1}          \fi
\ifx \showarticletitle \undefined \def \showarticletitle #1{#1}   \fi
\ifx \showURL      \undefined \def \showURL       {\relax}        \fi
\providecommand\bibfield[2]{#2}
\providecommand\bibinfo[2]{#2}
\providecommand\natexlab[1]{#1}
\providecommand\showeprint[2][]{arXiv:#2}

\bibitem[Ban et~al\mbox{.}(2025)]%
        {ban2025understanding}
\bibfield{author}{\bibinfo{person}{Yuanhao Ban}, \bibinfo{person}{Ruochen Wang}, \bibinfo{person}{Tianyi Zhou}, \bibinfo{person}{Minhao Cheng}, \bibinfo{person}{Boqing Gong}, {and} \bibinfo{person}{Cho-Jui Hsieh}.} \bibinfo{year}{2025}\natexlab{}.
\newblock \showarticletitle{Understanding the Impact of Negative Prompts: When and How Do They Take Effect?}. In \bibinfo{booktitle}{\emph{European Conference on Computer Vision}}. Springer, \bibinfo{pages}{190--206}.
\newblock


\bibitem[Chaliasos et~al\mbox{.}(2024)]%
        {chaliasos2024smart}
\bibfield{author}{\bibinfo{person}{Stefanos Chaliasos}, \bibinfo{person}{Marcos~Antonios Charalambous}, \bibinfo{person}{Liyi Zhou}, \bibinfo{person}{Rafaila Galanopoulou}, \bibinfo{person}{Arthur Gervais}, \bibinfo{person}{Dimitris Mitropoulos}, {and} \bibinfo{person}{Benjamin Livshits}.} \bibinfo{year}{2024}\natexlab{}.
\newblock \showarticletitle{Smart Contract and DeFi Security Tools: Do They Meet the Needs of Practitioners?}. In \bibinfo{booktitle}{\emph{Proceedings of the 46th IEEE/ACM International Conference on Software Engineering}}. \bibinfo{pages}{1--13}.
\newblock


\bibitem[Code4rena(2023)]%
        {code4rena}
\bibfield{author}{\bibinfo{person}{Code4rena}.} \bibinfo{year}{2023}\natexlab{}.
\newblock \bibinfo{booktitle}{\emph{Code4rena Contest Platform}}.
\newblock
\urldef\tempurl%
\url{https://code4rena.com/}
\showURL{%
\tempurl}
\newblock
\shownote{Accessed: 2023-04-16}.


\bibitem[Consensys(2024)]%
        {consensys2024mythril}
\bibfield{author}{\bibinfo{person}{Consensys}.} \bibinfo{year}{2024}\natexlab{}.
\newblock \bibinfo{booktitle}{\emph{Mythril: Security analysis tool for EVM bytecode}}.
\newblock
\urldef\tempurl%
\url{https://github.com/Consensys/mythril}
\showURL{%
\tempurl}
\newblock
\shownote{Accessed: 2024-06-06}.


\bibitem[David et~al\mbox{.}(2023)]%
        {david2023you}
\bibfield{author}{\bibinfo{person}{Isaac David}, \bibinfo{person}{Liyi Zhou}, \bibinfo{person}{Kaihua Qin}, \bibinfo{person}{Dawn Song}, \bibinfo{person}{Lorenzo Cavallaro}, {and} \bibinfo{person}{Arthur Gervais}.} \bibinfo{year}{2023}\natexlab{}.
\newblock \showarticletitle{Do you still need a manual smart contract audit?}
\newblock \bibinfo{journal}{\emph{arXiv preprint arXiv:2306.12338}} (\bibinfo{year}{2023}).
\newblock


\bibitem[Deng et~al\mbox{.}(2024)]%
        {deng2024safeguarding}
\bibfield{author}{\bibinfo{person}{Xun Deng}, \bibinfo{person}{Sidi~Mohamed Beillahi}, \bibinfo{person}{Cyrus Minwalla}, \bibinfo{person}{Han Du}, \bibinfo{person}{Andreas Veneris}, {and} \bibinfo{person}{Fan Long}.} \bibinfo{year}{2024}\natexlab{}.
\newblock \showarticletitle{Safeguarding DeFi Smart Contracts against Oracle Deviations}. In \bibinfo{booktitle}{\emph{Proceedings of the IEEE/ACM 46th International Conference on Software Engineering}}. \bibinfo{pages}{1--12}.
\newblock


\bibitem[Dominik(2025)]%
        {offchainoracle}
\bibfield{author}{\bibinfo{person}{Dominik}.} \bibinfo{year}{2025}\natexlab{}.
\newblock \bibinfo{booktitle}{\emph{Smart Contract Security Field Guide}}.
\newblock
\urldef\tempurl%
\url{https://scsfg.io/hackers/oracle-manipulation/#off-chain-infrastructure}
\showURL{%
\tempurl}
\newblock
\shownote{Accessed: 2025-01-06}.


\bibitem[Feist et~al\mbox{.}(2019)]%
        {feist2019slither}
\bibfield{author}{\bibinfo{person}{Josselin Feist}, \bibinfo{person}{Gustavo Grieco}, {and} \bibinfo{person}{Alex Groce}.} \bibinfo{year}{2019}\natexlab{}.
\newblock \showarticletitle{Slither: a static analysis framework for smart contracts}. In \bibinfo{booktitle}{\emph{2019 IEEE/ACM 2nd International Workshop on Emerging Trends in Software Engineering for Blockchain (WETSEB)}}. IEEE, \bibinfo{pages}{8--15}.
\newblock


\bibitem[Fu et~al\mbox{.}(2022)]%
        {fu2022complexity}
\bibfield{author}{\bibinfo{person}{Yao Fu}, \bibinfo{person}{Hao Peng}, \bibinfo{person}{Ashish Sabharwal}, \bibinfo{person}{Peter Clark}, {and} \bibinfo{person}{Tushar Khot}.} \bibinfo{year}{2022}\natexlab{}.
\newblock \showarticletitle{Complexity-based prompting for multi-step reasoning}.
\newblock \bibinfo{journal}{\emph{arXiv preprint arXiv:2210.00720}} (\bibinfo{year}{2022}).
\newblock


\bibitem[Gao et~al\mbox{.}(2021)]%
        {gao2021sverify}
\bibfield{author}{\bibinfo{person}{Bo Gao}, \bibinfo{person}{Ling Shi}, \bibinfo{person}{Jiaying Li}, \bibinfo{person}{Jialiang Chang}, \bibinfo{person}{Jun Sun}, {and} \bibinfo{person}{Zijiang Yang}.} \bibinfo{year}{2021}\natexlab{}.
\newblock \showarticletitle{sVerify: Verifying Smart Contracts Through Lazy Annotation and Learning}. In \bibinfo{booktitle}{\emph{Leveraging Applications of Formal Methods, Verification and Validation: 10th International Symposium on Leveraging Applications of Formal Methods, ISoLA 2021, Rhodes, Greece, October 17--29, 2021, Proceedings 10}}. Springer, \bibinfo{pages}{453--469}.
\newblock


\bibitem[Gao et~al\mbox{.}(2024)]%
        {gao2024unveiling}
\bibfield{author}{\bibinfo{person}{Bo Gao}, \bibinfo{person}{Qingsong Wei}, \bibinfo{person}{Yong Liu}, {and} \bibinfo{person}{Rick Siow~Mong Goh}.} \bibinfo{year}{2024}\natexlab{}.
\newblock \showarticletitle{Unveiling the Potential of ChatGPT in Detecting Machine Unauditable Bugs in Smart Contracts: A Preliminary Evaluation and Categorization}. In \bibinfo{booktitle}{\emph{2024 IEEE Conference on Artificial Intelligence (CAI)}}. IEEE, \bibinfo{pages}{1481--1486}.
\newblock


\bibitem[Gao et~al\mbox{.}(2023)]%
        {gao2023retrieval}
\bibfield{author}{\bibinfo{person}{Yunfan Gao}, \bibinfo{person}{Yun Xiong}, \bibinfo{person}{Xinyu Gao}, \bibinfo{person}{Kangxiang Jia}, \bibinfo{person}{Jinliu Pan}, \bibinfo{person}{Yuxi Bi}, \bibinfo{person}{Yi Dai}, \bibinfo{person}{Jiawei Sun}, {and} \bibinfo{person}{Haofen Wang}.} \bibinfo{year}{2023}\natexlab{}.
\newblock \showarticletitle{Retrieval-augmented generation for large language models: A survey}.
\newblock \bibinfo{journal}{\emph{arXiv preprint arXiv:2312.10997}} (\bibinfo{year}{2023}).
\newblock


\bibitem[Hu et~al\mbox{.}(2023)]%
        {hu2023large}
\bibfield{author}{\bibinfo{person}{Sihao Hu}, \bibinfo{person}{Tiansheng Huang}, \bibinfo{person}{Fatih {\.I}lhan}, \bibinfo{person}{Selim~Furkan Tekin}, {and} \bibinfo{person}{Ling Liu}.} \bibinfo{year}{2023}\natexlab{}.
\newblock \showarticletitle{Large language model-powered smart contract vulnerability detection: New perspectives}.
\newblock \bibinfo{journal}{\emph{arXiv preprint arXiv:2310.01152}} (\bibinfo{year}{2023}).
\newblock


\bibitem[Jiang et~al\mbox{.}(2018)]%
        {jiang2018contractfuzzer}
\bibfield{author}{\bibinfo{person}{Bo Jiang}, \bibinfo{person}{Ye Liu}, {and} \bibinfo{person}{Wing~Kwong Chan}.} \bibinfo{year}{2018}\natexlab{}.
\newblock \showarticletitle{Contractfuzzer: Fuzzing smart contracts for vulnerability detection}. In \bibinfo{booktitle}{\emph{Proceedings of the 33rd ACM/IEEE international conference on automated software engineering}}. \bibinfo{pages}{259--269}.
\newblock


\bibitem[Kojima et~al\mbox{.}(2022)]%
        {kojima2022large}
\bibfield{author}{\bibinfo{person}{Takeshi Kojima}, \bibinfo{person}{Shixiang~Shane Gu}, \bibinfo{person}{Machel Reid}, \bibinfo{person}{Yutaka Matsuo}, {and} \bibinfo{person}{Yusuke Iwasawa}.} \bibinfo{year}{2022}\natexlab{}.
\newblock \showarticletitle{Large language models are zero-shot reasoners}.
\newblock \bibinfo{journal}{\emph{Advances in neural information processing systems}}  \bibinfo{volume}{35} (\bibinfo{year}{2022}), \bibinfo{pages}{22199--22213}.
\newblock


\bibitem[Kong et~al\mbox{.}(2023)]%
        {kong2023defitainter}
\bibfield{author}{\bibinfo{person}{Queping Kong}, \bibinfo{person}{Jiachi Chen}, \bibinfo{person}{Yanlin Wang}, \bibinfo{person}{Zigui Jiang}, {and} \bibinfo{person}{Zibin Zheng}.} \bibinfo{year}{2023}\natexlab{}.
\newblock \showarticletitle{Defitainter: Detecting price manipulation vulnerabilities in defi protocols}. In \bibinfo{booktitle}{\emph{Proceedings of the 32nd ACM SIGSOFT International Symposium on Software Testing and Analysis}}. \bibinfo{pages}{1144--1156}.
\newblock


\bibitem[Labadie(2022)]%
        {labadie2022impermanent}
\bibfield{author}{\bibinfo{person}{Mauricio Labadie}.} \bibinfo{year}{2022}\natexlab{}.
\newblock \showarticletitle{Impermanent loss and slippage in Automated Market Makers (AMMs) with constant-product formula}.
\newblock \bibinfo{journal}{\emph{Available at SSRN 4053924}} (\bibinfo{year}{2022}).
\newblock


\bibitem[Li et~al\mbox{.}(2023)]%
        {li2023adapting}
\bibfield{author}{\bibinfo{person}{Qingyao Li}, \bibinfo{person}{Lingyue Fu}, \bibinfo{person}{Weiming Zhang}, \bibinfo{person}{Xianyu Chen}, \bibinfo{person}{Jingwei Yu}, \bibinfo{person}{Wei Xia}, \bibinfo{person}{Weinan Zhang}, \bibinfo{person}{Ruiming Tang}, {and} \bibinfo{person}{Yong Yu}.} \bibinfo{year}{2023}\natexlab{}.
\newblock \showarticletitle{Adapting large language models for education: Foundational capabilities, potentials, and challenges}.
\newblock \bibinfo{journal}{\emph{arXiv preprint arXiv:2401.08664}} (\bibinfo{year}{2023}).
\newblock


\bibitem[Liu et~al\mbox{.}(2021)]%
        {liu2021combining}
\bibfield{author}{\bibinfo{person}{Zhenguang Liu}, \bibinfo{person}{Peng Qian}, \bibinfo{person}{Xiaoyang Wang}, \bibinfo{person}{Yuan Zhuang}, \bibinfo{person}{Lin Qiu}, {and} \bibinfo{person}{Xun Wang}.} \bibinfo{year}{2021}\natexlab{}.
\newblock \showarticletitle{Combining graph neural networks with expert knowledge for smart contract vulnerability detection}.
\newblock \bibinfo{journal}{\emph{IEEE Transactions on Knowledge and Data Engineering}} \bibinfo{volume}{35}, \bibinfo{number}{2} (\bibinfo{year}{2021}), \bibinfo{pages}{1296--1310}.
\newblock


\bibitem[Longpre et~al\mbox{.}(2023)]%
        {longpre2023flan}
\bibfield{author}{\bibinfo{person}{Shayne Longpre}, \bibinfo{person}{Le Hou}, \bibinfo{person}{Tu Vu}, \bibinfo{person}{Albert Webson}, \bibinfo{person}{Hyung~Won Chung}, \bibinfo{person}{Yi Tay}, \bibinfo{person}{Denny Zhou}, \bibinfo{person}{Quoc~V Le}, \bibinfo{person}{Barret Zoph}, \bibinfo{person}{Jason Wei}, {et~al\mbox{.}}} \bibinfo{year}{2023}\natexlab{}.
\newblock \showarticletitle{The flan collection: Designing data and methods for effective instruction tuning}. In \bibinfo{booktitle}{\emph{International Conference on Machine Learning}}. PMLR, \bibinfo{pages}{22631--22648}.
\newblock


\bibitem[Luu et~al\mbox{.}(2016)]%
        {luu2016oyente}
\bibfield{author}{\bibinfo{person}{Loi Luu}, \bibinfo{person}{Duc-Hiep Chu}, \bibinfo{person}{Hrishi Olickel}, \bibinfo{person}{Prateek Saxena}, {and} \bibinfo{person}{Aquinas Hobor}.} \bibinfo{year}{2016}\natexlab{}.
\newblock \showarticletitle{Making smart contracts smarter}. In \bibinfo{booktitle}{\emph{Proceedings of the 2016 ACM SIGSAC conference on computer and communications security}}. \bibinfo{pages}{254--269}.
\newblock


\bibitem[Mo et~al\mbox{.}(2023)]%
        {mo2023toward}
\bibfield{author}{\bibinfo{person}{Yifan Mo}, \bibinfo{person}{Jiachi Chen}, \bibinfo{person}{Yanlin Wang}, {and} \bibinfo{person}{Zibin Zheng}.} \bibinfo{year}{2023}\natexlab{}.
\newblock \showarticletitle{Toward automated detecting unanticipated price feed in smart contract}. In \bibinfo{booktitle}{\emph{Proceedings of the 32nd ACM SIGSOFT International Symposium on Software Testing and Analysis}}. \bibinfo{pages}{1257--1268}.
\newblock


\bibitem[Mossberg et~al\mbox{.}(2019)]%
        {mossberg2019manticore}
\bibfield{author}{\bibinfo{person}{Mark Mossberg}, \bibinfo{person}{Felipe Manzano}, \bibinfo{person}{Eric Hennenfent}, \bibinfo{person}{Alex Groce}, \bibinfo{person}{Gustavo Grieco}, \bibinfo{person}{Josselin Feist}, \bibinfo{person}{Trent Brunson}, {and} \bibinfo{person}{Artem Dinaburg}.} \bibinfo{year}{2019}\natexlab{}.
\newblock \showarticletitle{Manticore: A user-friendly symbolic execution framework for binaries and smart contracts}. In \bibinfo{booktitle}{\emph{2019 34th IEEE/ACM International Conference on Automated Software Engineering (ASE)}}. IEEE, \bibinfo{pages}{1186--1189}.
\newblock


\bibitem[Mueller et~al\mbox{.}(2024)]%
        {mueller2024mythril}
\bibfield{author}{\bibinfo{person}{Bernhard Mueller}, \bibinfo{person}{Nikhil Parasaram}, \bibinfo{person}{Joran Honig}, {and} \bibinfo{person}{Dominik Muhs}.} \bibinfo{year}{2024}\natexlab{}.
\newblock \bibinfo{title}{Mythril}.
\newblock \bibinfo{howpublished}{\url{https://github.com/Consensys/mythril}}.
\newblock
\newblock
\shownote{Accessed: 2024-05-23}.


\bibitem[Nguyen et~al\mbox{.}(2020)]%
        {nguyen2020sfuzz}
\bibfield{author}{\bibinfo{person}{Tai~D Nguyen}, \bibinfo{person}{Long~H Pham}, \bibinfo{person}{Jun Sun}, \bibinfo{person}{Yun Lin}, {and} \bibinfo{person}{Quang~Tran Minh}.} \bibinfo{year}{2020}\natexlab{}.
\newblock \showarticletitle{sfuzz: An efficient adaptive fuzzer for solidity smart contracts}. In \bibinfo{booktitle}{\emph{Proceedings of the ACM/IEEE 42nd International Conference on Software Engineering}}. \bibinfo{pages}{778--788}.
\newblock


\bibitem[Openai(2024)]%
        {openaiparams}
\bibfield{author}{\bibinfo{person}{Openai}.} \bibinfo{year}{2024}\natexlab{}.
\newblock \bibinfo{booktitle}{\emph{Openai API References}}.
\newblock
\urldef\tempurl%
\url{https://platform.openai.com/docs/api-reference/chat/create}
\showURL{%
\tempurl}
\newblock
\shownote{Accessed: 2024-05-26}.


\bibitem[Protofire(2024)]%
        {protofire2024solhint}
\bibfield{author}{\bibinfo{person}{Protofire}.} \bibinfo{year}{2024}\natexlab{}.
\newblock \bibinfo{booktitle}{\emph{Solhint: provide a linting utility for Solidity code}}.
\newblock
\urldef\tempurl%
\url{https://github.com/protofire/solhint}
\showURL{%
\tempurl}
\newblock
\shownote{Accessed: 2024-06-06}.


\bibitem[So et~al\mbox{.}(2020)]%
        {so2020verismart}
\bibfield{author}{\bibinfo{person}{Sunbeom So}, \bibinfo{person}{Myungho Lee}, \bibinfo{person}{Jisu Park}, \bibinfo{person}{Heejo Lee}, {and} \bibinfo{person}{Hakjoo Oh}.} \bibinfo{year}{2020}\natexlab{}.
\newblock \showarticletitle{VeriSmart: A highly precise safety verifier for Ethereum smart contracts}. In \bibinfo{booktitle}{\emph{2020 IEEE Symposium on Security and Privacy (SP)}}. IEEE, \bibinfo{pages}{1678--1694}.
\newblock


\bibitem[Sun et~al\mbox{.}(2023)]%
        {sun2023gptscan}
\bibfield{author}{\bibinfo{person}{Yuqiang Sun}, \bibinfo{person}{Daoyuan Wu}, \bibinfo{person}{Yue Xue}, \bibinfo{person}{Han Liu}, \bibinfo{person}{Haijun Wang}, \bibinfo{person}{Zhengzi Xu}, \bibinfo{person}{Xiaofei Xie}, {and} \bibinfo{person}{Yang Liu}.} \bibinfo{year}{2023}\natexlab{}.
\newblock \showarticletitle{When gpt meets program analysis: Towards intelligent detection of smart contract logic vulnerabilities in gptscan}.
\newblock \bibinfo{journal}{\emph{arXiv preprint arXiv:2308.03314}} (\bibinfo{year}{2023}).
\newblock


\bibitem[Tikhomirov et~al\mbox{.}(2018)]%
        {tikhomirov2018smartcheck}
\bibfield{author}{\bibinfo{person}{Sergei Tikhomirov}, \bibinfo{person}{Ekaterina Voskresenskaya}, \bibinfo{person}{Ivan Ivanitskiy}, \bibinfo{person}{Ramil Takhaviev}, \bibinfo{person}{Evgeny Marchenko}, {and} \bibinfo{person}{Yaroslav Alexandrov}.} \bibinfo{year}{2018}\natexlab{}.
\newblock \showarticletitle{Smartcheck: Static analysis of ethereum smart contracts}. In \bibinfo{booktitle}{\emph{Proceedings of the 1st international workshop on emerging trends in software engineering for blockchain}}. \bibinfo{pages}{9--16}.
\newblock


\bibitem[Torres et~al\mbox{.}(2021)]%
        {torres2021confuzzius}
\bibfield{author}{\bibinfo{person}{Christof~Ferreira Torres}, \bibinfo{person}{Antonio~Ken Iannillo}, \bibinfo{person}{Arthur Gervais}, {and} \bibinfo{person}{Radu State}.} \bibinfo{year}{2021}\natexlab{}.
\newblock \showarticletitle{Confuzzius: A data dependency-aware hybrid fuzzer for smart contracts}. In \bibinfo{booktitle}{\emph{2021 IEEE European Symposium on Security and Privacy (EuroS\&P)}}. IEEE, \bibinfo{pages}{103--119}.
\newblock


\bibitem[Wang et~al\mbox{.}(2021)]%
        {wang2021promutator}
\bibfield{author}{\bibinfo{person}{Shih-Hung Wang}, \bibinfo{person}{Chia-Chien Wu}, \bibinfo{person}{Yu-Chuan Liang}, \bibinfo{person}{Li-Hsun Hsieh}, {and} \bibinfo{person}{Hsu-Chun Hsiao}.} \bibinfo{year}{2021}\natexlab{}.
\newblock \showarticletitle{ProMutator: Detecting vulnerable price oracles in DeFi by mutated transactions}. In \bibinfo{booktitle}{\emph{2021 IEEE European Symposium on Security and Privacy Workshops (EuroS\&PW)}}. IEEE, \bibinfo{pages}{380--385}.
\newblock


\bibitem[Wang et~al\mbox{.}(2020)]%
        {wang2020contractward}
\bibfield{author}{\bibinfo{person}{Wei Wang}, \bibinfo{person}{Jingjing Song}, \bibinfo{person}{Guangquan Xu}, \bibinfo{person}{Yidong Li}, \bibinfo{person}{Hao Wang}, {and} \bibinfo{person}{Chunhua Su}.} \bibinfo{year}{2020}\natexlab{}.
\newblock \showarticletitle{Contractward: Automated vulnerability detection models for ethereum smart contracts}.
\newblock \bibinfo{journal}{\emph{IEEE Transactions on Network Science and Engineering}} \bibinfo{volume}{8}, \bibinfo{number}{2} (\bibinfo{year}{2020}), \bibinfo{pages}{1133--1144}.
\newblock


\bibitem[Wei et~al\mbox{.}(2022)]%
        {wei2022chain}
\bibfield{author}{\bibinfo{person}{Jason Wei}, \bibinfo{person}{Xuezhi Wang}, \bibinfo{person}{Dale Schuurmans}, \bibinfo{person}{Maarten Bosma}, \bibinfo{person}{Fei Xia}, \bibinfo{person}{Ed Chi}, \bibinfo{person}{Quoc~V Le}, \bibinfo{person}{Denny Zhou}, {et~al\mbox{.}}} \bibinfo{year}{2022}\natexlab{}.
\newblock \showarticletitle{Chain-of-thought prompting elicits reasoning in large language models}.
\newblock \bibinfo{journal}{\emph{Advances in Neural Information Processing Systems}}  \bibinfo{volume}{35} (\bibinfo{year}{2022}), \bibinfo{pages}{24824--24837}.
\newblock


\bibitem[Wen et~al\mbox{.}(2024)]%
        {wen2024foray}
\bibfield{author}{\bibinfo{person}{Hongbo Wen}, \bibinfo{person}{Hanzhi Liu}, \bibinfo{person}{Jiaxin Song}, \bibinfo{person}{Yanju Chen}, \bibinfo{person}{Wenbo Guo}, {and} \bibinfo{person}{Yu Feng}.} \bibinfo{year}{2024}\natexlab{}.
\newblock \showarticletitle{FORAY: Towards Effective Attack Synthesis against Deep Logical Vulnerabilities in DeFi Protocols}.
\newblock \bibinfo{journal}{\emph{arXiv preprint arXiv:2407.06348}} (\bibinfo{year}{2024}).
\newblock


\bibitem[Wood et~al\mbox{.}(2014)]%
        {wood2014ethereum}
\bibfield{author}{\bibinfo{person}{Gavin Wood} {et~al\mbox{.}}} \bibinfo{year}{2014}\natexlab{}.
\newblock \showarticletitle{Ethereum: A secure decentralised generalised transaction ledger}.
\newblock \bibinfo{journal}{\emph{Ethereum project yellow paper}} \bibinfo{volume}{151}, \bibinfo{number}{2014} (\bibinfo{year}{2014}), \bibinfo{pages}{1--32}.
\newblock


\bibitem[Wu et~al\mbox{.}(2021)]%
        {wu2021defiranger}
\bibfield{author}{\bibinfo{person}{Siwei Wu}, \bibinfo{person}{Dabao Wang}, \bibinfo{person}{Jianting He}, \bibinfo{person}{Yajin Zhou}, \bibinfo{person}{Lei Wu}, \bibinfo{person}{Xingliang Yuan}, \bibinfo{person}{Qinming He}, {and} \bibinfo{person}{Kui Ren}.} \bibinfo{year}{2021}\natexlab{}.
\newblock \showarticletitle{Defiranger: Detecting price manipulation attacks on defi applications}.
\newblock \bibinfo{journal}{\emph{arXiv preprint arXiv:2104.15068}} (\bibinfo{year}{2021}).
\newblock


\bibitem[Xi et~al\mbox{.}(2024)]%
        {xi2024pomabuster}
\bibfield{author}{\bibinfo{person}{Rui Xi}, \bibinfo{person}{Zehua Wang}, {and} \bibinfo{person}{Karthik Pattabiraman}.} \bibinfo{year}{2024}\natexlab{}.
\newblock \showarticletitle{POMABuster: Detecting Price Oracle Manipulation Attacks in Decentralized Finance}. In \bibinfo{booktitle}{\emph{2024 IEEE Symposium on Security and Privacy (SP)}}. IEEE Computer Society, \bibinfo{pages}{240--240}.
\newblock


\bibitem[Zeng et~al\mbox{.}(2022)]%
        {zeng2022ethergis}
\bibfield{author}{\bibinfo{person}{Qingren Zeng}, \bibinfo{person}{Jiahao He}, \bibinfo{person}{Gansen Zhao}, \bibinfo{person}{Shuangyin Li}, \bibinfo{person}{Jingji Yang}, \bibinfo{person}{Hua Tang}, {and} \bibinfo{person}{Haoyu Luo}.} \bibinfo{year}{2022}\natexlab{}.
\newblock \showarticletitle{EtherGIS: A Vulnerability Detection Framework for Ethereum Smart Contracts Based on Graph Learning Features}. In \bibinfo{booktitle}{\emph{2022 IEEE 46th Annual Computers, Software, and Applications Conference (COMPSAC)}}. IEEE, \bibinfo{pages}{1742--1749}.
\newblock


\bibitem[Zhang et~al\mbox{.}(2023)]%
        {zhang2023demystifying}
\bibfield{author}{\bibinfo{person}{Zhuo Zhang}, \bibinfo{person}{Brian Zhang}, \bibinfo{person}{Wen Xu}, {and} \bibinfo{person}{Zhiqiang Lin}.} \bibinfo{year}{2023}\natexlab{}.
\newblock \showarticletitle{Demystifying exploitable bugs in smart contracts}. In \bibinfo{booktitle}{\emph{2023 IEEE/ACM 45th International Conference on Software Engineering (ICSE)}}. IEEE, \bibinfo{pages}{615--627}.
\newblock


\bibitem[Zheng et~al\mbox{.}(2023)]%
        {zheng2023helpful}
\bibfield{author}{\bibinfo{person}{Mingqian Zheng}, \bibinfo{person}{Jiaxin Pei}, {and} \bibinfo{person}{David Jurgens}.} \bibinfo{year}{2023}\natexlab{}.
\newblock \showarticletitle{Is" a helpful assistant" the best role for large language models? a systematic evaluation of social roles in system prompts}.
\newblock \bibinfo{journal}{\emph{arXiv preprint arXiv:2311.10054}}  \bibinfo{volume}{8} (\bibinfo{year}{2023}).
\newblock


\bibitem[Zhou et~al\mbox{.}(2022)]%
        {zhou2022least}
\bibfield{author}{\bibinfo{person}{Denny Zhou}, \bibinfo{person}{Nathanael Sch{\"a}rli}, \bibinfo{person}{Le Hou}, \bibinfo{person}{Jason Wei}, \bibinfo{person}{Nathan Scales}, \bibinfo{person}{Xuezhi Wang}, \bibinfo{person}{Dale Schuurmans}, \bibinfo{person}{Claire Cui}, \bibinfo{person}{Olivier Bousquet}, \bibinfo{person}{Quoc Le}, {et~al\mbox{.}}} \bibinfo{year}{2022}\natexlab{}.
\newblock \showarticletitle{Least-to-most prompting enables complex reasoning in large language models}.
\newblock \bibinfo{journal}{\emph{arXiv preprint arXiv:2205.10625}} (\bibinfo{year}{2022}).
\newblock


\bibitem[Zhou et~al\mbox{.}(2021)]%
        {zhou2021defiposer}
\bibfield{author}{\bibinfo{person}{Liyi Zhou}, \bibinfo{person}{Kaihua Qin}, \bibinfo{person}{Antoine Cully}, \bibinfo{person}{Benjamin Livshits}, {and} \bibinfo{person}{Arthur Gervais}.} \bibinfo{year}{2021}\natexlab{}.
\newblock \showarticletitle{On the just-in-time discovery of profit-generating transactions in defi protocols}. In \bibinfo{booktitle}{\emph{2021 IEEE Symposium on Security and Privacy (SP)}}. IEEE, \bibinfo{pages}{919--936}.
\newblock


\bibitem[Zhou et~al\mbox{.}(2023)]%
        {zhou2023sok}
\bibfield{author}{\bibinfo{person}{Liyi Zhou}, \bibinfo{person}{Xihan Xiong}, \bibinfo{person}{Jens Ernstberger}, \bibinfo{person}{Stefanos Chaliasos}, \bibinfo{person}{Zhipeng Wang}, \bibinfo{person}{Ye Wang}, \bibinfo{person}{Kaihua Qin}, \bibinfo{person}{Roger Wattenhofer}, \bibinfo{person}{Dawn Song}, {and} \bibinfo{person}{Arthur Gervais}.} \bibinfo{year}{2023}\natexlab{}.
\newblock \showarticletitle{Sok: Decentralized finance (defi) attacks}. In \bibinfo{booktitle}{\emph{2023 IEEE Symposium on Security and Privacy (SP)}}. IEEE, \bibinfo{pages}{2444--2461}.
\newblock


\end{thebibliography}
